\documentclass{emulateapj}

\newcommand{\her}{{\it Herschel}}

\newcommand{\water}   {H$_2$O}

\newcommand{\Tex}   {$T_\mathrm{ex}$}

\newcommand{\mum}   {$\mu$m}
\newcommand{\kms}   {km~s$^{-1}$}

\newcommand{\cmd}   {cm$^{-2}$}
\newcommand{\cmt}   {cm$^{-3}$}

\newcommand{\hho} {H$_2$O}
\newcommand{\hh}    {H$_2$}

\newcommand{\htop}	{H$_3$O$^+$}
\newcommand{\hdop}	{H$_2$O$^+$}
\newcommand{\ohp}		{OH$^+$}

\shorttitle{Submillimeter spectrum of NGC\,1068}
\shortauthors{Spinoglio et al.}

\usepackage{graphicx}
\usepackage{amsmath}

\begin{document}

% \DeclareGraphicsExtensions{.pdf,.gif,.jpg}

\title{Submillimetre line spectrum of the Seyfert galaxy NGC\,1068 from the \her -SPIRE Fourier Transform Spectrometer \footnotemark[$\star$]} 
\footnotetext[$\star$]{Herschel is an ESA space observatory with science instruments provided by European-led Principal Investigator consortia and with important participation from NASA.}

\author{
Luigi Spinoglio\altaffilmark{1}, 
Miguel Pereira-Santaella\altaffilmark{1}, 
Gemma Busquet\altaffilmark{1} , 
Maximilien R. P. Schirm\altaffilmark{2},  
Christine D. Wilson\altaffilmark{2} , 
Jason Glenn\altaffilmark{3}, 
Julia Kamenetzky\altaffilmark{3}, 
Naseem Rangwala\altaffilmark{3}, 
Philip R. Maloney\altaffilmark{3}, 
Tara J. Parkin\altaffilmark{2},  
George J. Bendo\altaffilmark{4}, 
Suzanne C. Madden\altaffilmark{5},
Mark G. Wolfire\altaffilmark{6},
Alessandro Boselli\altaffilmark{7},
Asantha Cooray\altaffilmark{8}, 
Mathew J. Page\altaffilmark{9}
}

\altaffiltext{1}{Istituto di Astrofisica e Planetologia Spaziali, INAF-IAPS, Via Fosso del Cavaliere 100, I-00133 Roma, Italy}
\email{luigi.spinoglio@iaps.inaf.it}
\altaffiltext{2}{Dept. of Physics \& Astronomy, McMaster University, Hamilton, Ontario, L8S 4M1, Canada}
\altaffiltext{3}{Center for Astrophysics and Space Astronomy, 389-UCB, University of Colorado, Boulder, CO, 80303}
\altaffiltext{4}{UK ALMA Regional Centre Node, Jordell Bank Center for Astrophysics, School of Physics and Astronomy, University of
Manchester, Oxford Road, Manchester M13 9PL, U.K.}
\altaffiltext{5}{CEA, Laboratoire AIM, Irfu/SAp, Orme des Merisiers, 91191 Gif-sur-Yvette, France}
\altaffiltext{6}{Astronomy Department, University of Maryland, College Park, MD 20742, USA}
\altaffiltext{7}{Laboratoire d'Astrophysique de Marseille - LAM, Universit\'e d'Aix-Marseille \& CNRS, UMR7326, 38 rue F. Joliot-Curie, 13388, Marseille Cedex 13}
\altaffiltext{8}{Center for Cosmology, Department of Physics and Astronomy, University of California, Irvine, CA 92697}
\altaffiltext{9}{Mullard Space Science Laboratory, University College London, Holmbury St Mary, Dorking, Surrey RH5 6NT}

\clearpage

\begin{abstract}

The first complete submillimetre spectrum (190-670$\mu$m) of the Seyfert 2 galaxy NGC\,1068 has been observed with the SPIRE Fourier Transform Spectrometer onboard the {\it Herschel} Space Observatory. The sequence of CO lines (J$_{up}$=4-13), lines from \water, the fundamental rotational transition of HF,  two o-H$_2$O$^{+}$ lines and one line each from CH$^{+}$ and OH$^{+}$ have been detected, together with the two [CI] lines and the [NII]205$\mu$m line. The observations in both single pointing mode with sparse image sampling and in mapping mode with full image sampling allow us to disentangle two molecular emission components, one due to the compact circum-nuclear disk (CND) and one from the extended region encompassing the star forming ring (SF-ring). Radiative transfer models show that the two CO components are characterized by density of $n({\rm H_2})$=$10^{4.5}$ and $10^{2.9}$ cm$^{-3}$ and temperature of $T_{\rm kin}$ =100K and  127K, respectively. The comparison of the CO line intensities with photodissociation region (PDR) and X-ray dominated region (XDR) models, together with other observational constraints, such as the observed CO surface brightness and the radiation field, indicate that the best explanation for the CO excitation of the CND is an XDR with density of n(H$_2$) $\sim$ 10$^4$ cm$^{-3}$ and X-ray flux of 9 erg~s$^{-1}$~cm$^{-2}$, consistent with illumination by the active galactic nucleus, while the CO lines in the SF-ring are better modeled by a PDR. 
The detected water transitions, together with those observed with the \her~ PACS Spectrometer, can be modeled by an LVG model with low temperature ($T_{\rm kin}\sim$ 40K) and high density ($n({\rm H_2})$ in the range $10^{6.7}$ --- $10^{7.9}$ cm$^{-3}$). The emission of \hdop\ and \ohp\ are in agreement with PDR models with cosmic ray ionization.  The diffuse ionized atomic component observed through the [NII]205$\mu$m line is consistent with previous photoionization models of the starburst.

\end{abstract}


\section{Introduction}
\label{intro}

NGC\,1068 (Messier 77) is a nearby ($cz$ = 1137 km s$^{-1}$) and bright (L$_{IR}$=L$_{8-1000\mu m}$ $\sim$2 $\times$ 10$^{11}$ L$_{\odot}$, \citealt{bla97}) 
Seyfert galaxy, often considered as the prototypical Seyfert type 2 galaxy.
However, since the discovery of the broad permitted lines in the polarized optical spectrum 
of NGC\,1068 \citep{anto85},
it has become clear that this galaxy was in reality a {\it hidden broad line region} galaxy. 
The general distinction between the two types of Seyfert galaxies might be due only to
orientation effects, according to the so called Unification model \citep{anto93}. 

Being the strongest nearby 
Seyfert 2 galaxy, it has been observed extensively over the whole electromagnetic spectrum. 
Molecular (CO and HCN) 
observations have shown a prominent starburst ring (hereafter SF-ring) at a radius of 1.0-1.5 kpc and a central circum-nuclear disk 
(hereafter CND) with a diameter of $d$ $\sim$ 300 pc \citep{tac94,sch00}. 
Near-IR observations \citep{sco88,thr89} have clearly revealed a 2.3 kpc stellar bar.
The compact ($\sim$1 pc) hot dust source in the nucleus of NGC\,1068, 
measured using near-infrared speckle imaging and 
integral field spectroscopy \citep{tha97}, is probably heated by the AGN's strong radiation 
field and possibly associated with the postulated dense circum-nuclear torus. 
A jet was observed from centimeter to millimeter
wavelengths extending out to several kiloparsecs from the center \citep[e.g.,][]{kri06,gal04}.
Mid-IR observations revealed hot and ionized gas biconically following the path of the radio jet 
\citep[e.g.,][and references therein]{mul09}
and indicating the existence of a parsec-scale warm dust torus \citep{jaf04}. 
Recent interferometric observations of CO (3-2) and CO (1-0) show that the CO (3-2) emission peaks in the central
region within $\sim$ 5 $\arcsec$ from the nucleus, while the CO(1-0) emission is mainly located along the spiral arms \citep{tsa12}.

Spectroscopic coverage of NGC\,1068  has been extensive.
In the mid-IR to far-IR wavelength range, spectra have been measured by the ISO \citep{kes96} SWS \citep{deg96}  and LWS  \citep{cle96} spectrometers 
\citep[][respectively]{lut00,spi05}, 
covering the 2.4-45~$\mu$m and 43-197~$\mu$m spectral ranges. 
At millimeter wavelengths it was recently observed from the ground in the
190-307 GHz (976-1578~$\mu$m) range \citep{kam11}.
The submillimeter waveband 
is one of the few spectral regions that have not been so far explored; the new
observations made by the Spectral and Photometric Imaging Receiver (SPIRE) \citep{gri10} Fourier Transform Spectrometer 
(FTS) \citep{nay10a}, onboard the \her~Space Observatory \citep{pil10}, 
covering the spectral range 
from 190~$\mu$m to 670~$\mu$m, fill most of this gap with the
first complete submillimeter spectrum of NGC\,1068.

Submillimeter  spectral measurements are of particular interest as NGC\,1068 is a prime
candidate to study the effects of the AGN onto the circum-nuclear material and the surrounding disk.
In particular, the study of the excitation conditions and chemistry of the CND appear already 
from the existing ground-based molecular line observations to be
very different from the starburst galaxies environments. The peculiar line ratios of different
molecular transitions, mostly HCN, HCO$^+$, and $^{12}$CO, led to the suggestion that the CND of 
NGC\,1068 harbors a giant X-ray-dominated region \citep[XDR, e.g.,][]{rot91, mal96, use04, koh08}. 
The HCN and HCO$^+$ molecular line studies of \citet{kri08,kri11} 
confirm an increased abundance of HCN and/or increased kinetic temperatures.
CO lines can also discriminate between  ``classical'' photodissociation regions (PDRs) and  
X-ray dominated regions (XDRs) \citep[e.g.][]{mei05}.
Using the intermediate J rotational lines from the CO molecule, from J$_{up}$=4 to J$_{up}$=13,  
which can be observed with the SPIRE FTS, 
we want to test if indeed in NGC\,1068 an XDR is needed to explain the spectral line energy distribution 
originating from the CND. 
The case of the ultraluminous IR galaxy Mrk231 already demonstrated that
the \her -SPIRE data are indeed able to discriminate between the two emission mechanisms and therefore detect
the effects of the AGN \citep{wer10}.
\her -PACS observations of the high-J CO lines (J$_{up} \ge 14$) detected in NGC\,1068 
have been recently presented by \citet{hai12}. The two components, at high and medium excitation, needed to explain 
the observed CO lines arising from the central 10$\arcsec$ region from 
J$_{up}$= 14 to J$_{up}$= 24 can be excited by X-ray or shock heating, while far-UV heating is unlikely.

The \her -SPIRE spectroscopic 
observations of NGC\,1068 presented here have been collected under the guaranteed time key project 
``Physical Processes in the Interstellar Medium of Very Nearby Galaxies'' (PI: Christine Wilson). Within the same observational program, 
SPIRE and Photodetector Array Camera and Spectrometer (PACS) \citep{pog10} photometric images 
have been collected, which will be presented in a forthcoming paper with the detailed analysis of the continuum emission 
(Spinoglio et al 2012, in prep.).

\section{Observations}
\label{sec:observations}

\subsection{SPIRE Spectroscopy}
%\subsubsection{Deep spectrum}

\begin{figure}[hb!]
\includegraphics[width=\columnwidth]{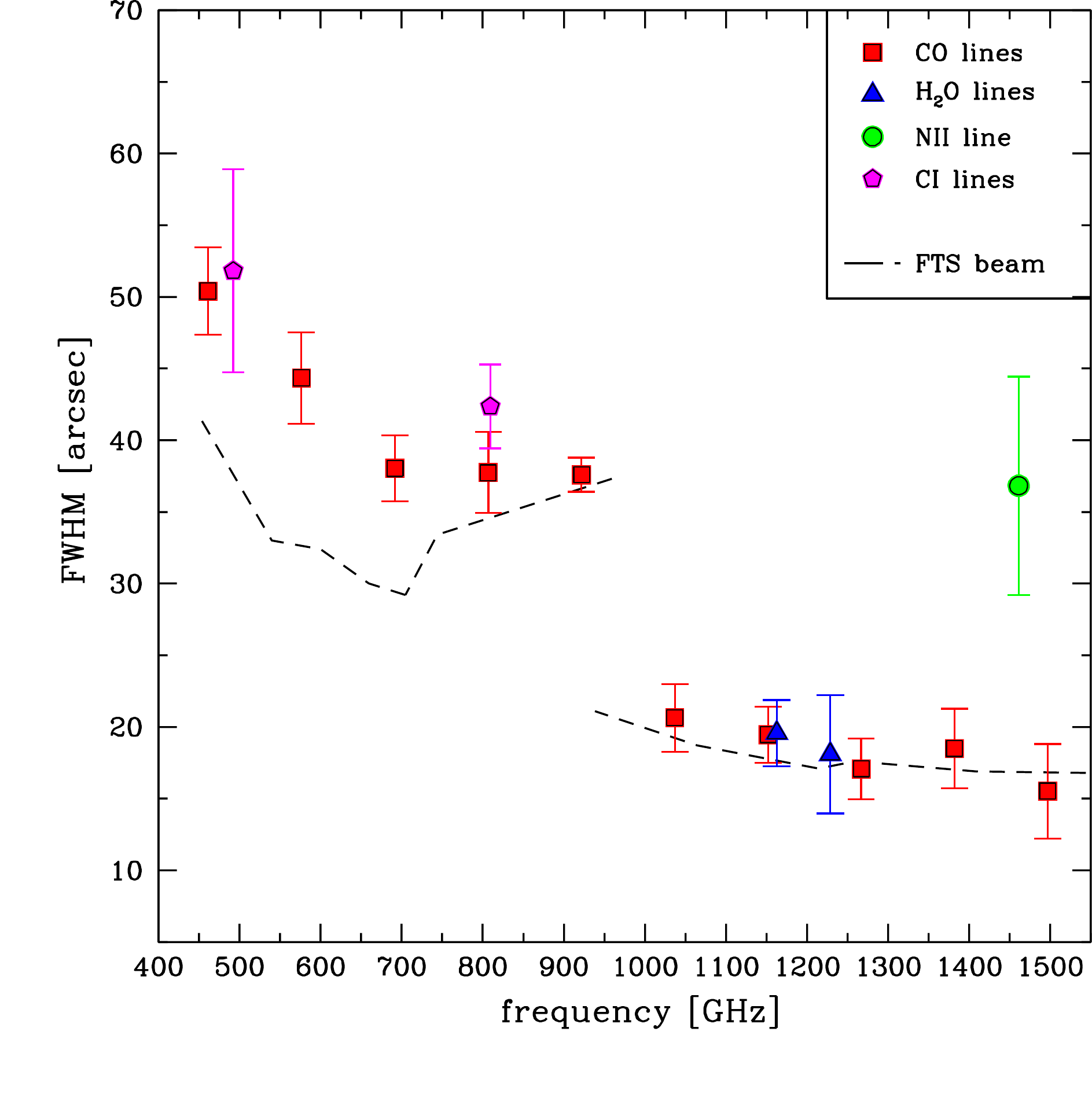}
\caption{Observed source sizes, as measured from the mapping observations of the brightest SPIRE lines (FWHM in arcsec, with 1$\sigma$ error bars) as a function of frequency (colored points) compared with the SPIRE-FTS beam (broken lines). 
All source sizes in the SSW, except for [NII]205~$\mu$m, are consistent with the beam width at that frequency.
In contrast, the source sizes in the SLW are more extended than the beam. }
\label{lines_fwhm}
\end{figure}

\begin{figure*}[ht!]
 \includegraphics[width=\columnwidth]{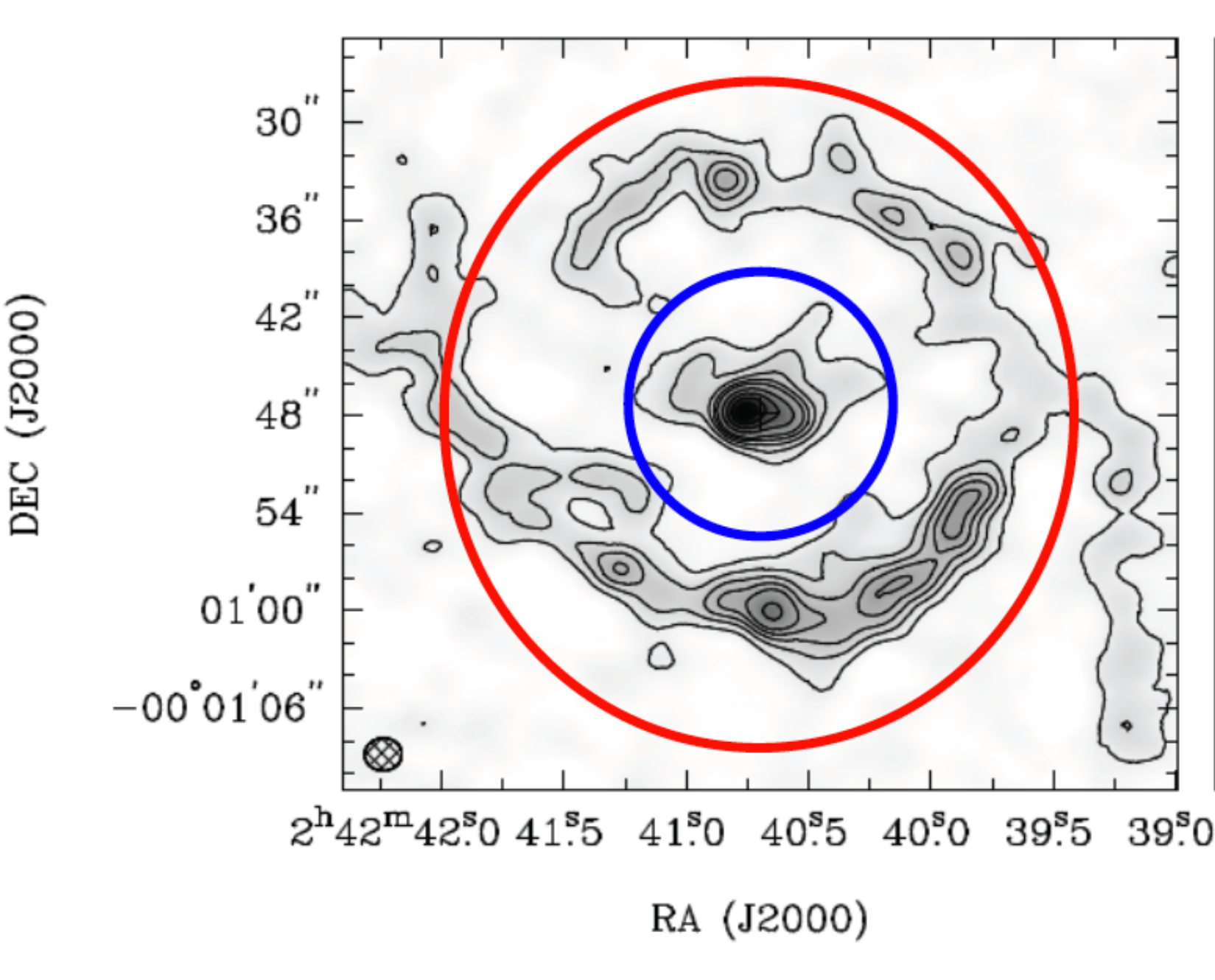}\includegraphics[width=\columnwidth]{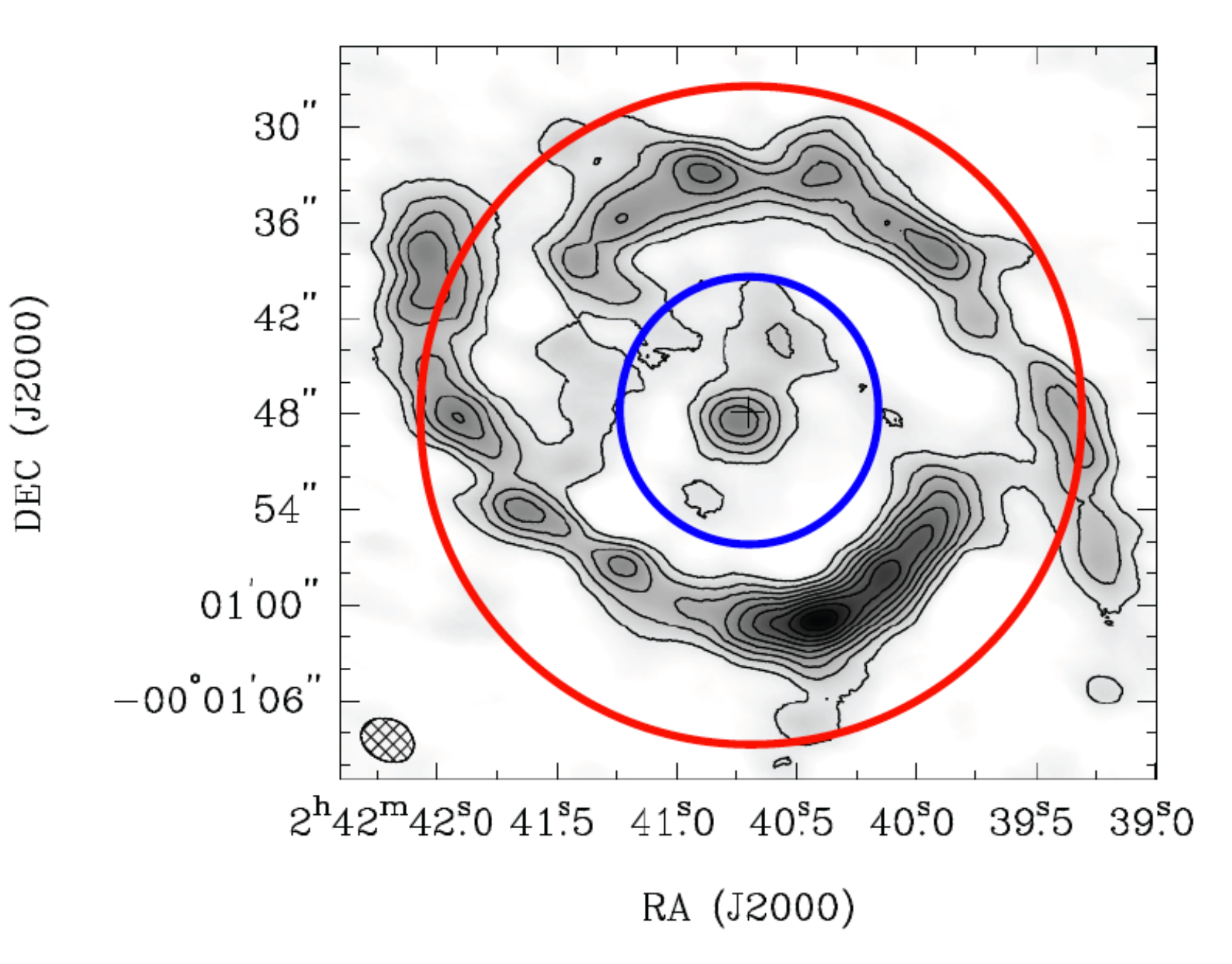}
\caption{
{\it Left:} CO(3-2) integrated intensity  map of the central region of NGC 1068 from \citet{tsa12}.
{\it Right:} CO(1-0) integrated intensity  map of the central region of NGC 1068 from \citet{tsa12}. 
The minimum SSW (\emph{blue}) and the maximum SLW (\emph{red}) SPIRE FTS beam sizes of 
17$\arcsec$ and 42$\arcsec$, respectively, are shown in the figures
as concentric circles. }
\label{CO_maps}
\end{figure*}

NGC\,1068 was observed with the SPIRE-FTS \citep{gri10} onboard the {\it Herschel} Space Observatory \citep{pil10} 
both in the single pointing mode with sparse image sampling and in the mapping mode with full image sampling. 
The FTS has two detectors arrays called the spectrometer long wave (SLW, in the range of 303-671~$\mu$m) 
and the spectrometer short wave (SSW, in the range 194-313~$\mu$m), with a small (10 $\mu$m) overlap in wavelength. 
The spectral resolution of the FTS ranges from about 280 km s$^{-1}$ to 950 km s$^{-1}$ 
in the high resolution mode, moving in wavelength from 194 $\mu$m to 671 $\mu$m.
The two SPIRE FTS observations of NGC\,1068 were collected on Operational Day 626 (29 January 2011); the fully sampled 
map has Observation ID 1342213444 and the single pointing "deep" spectrum has Observation ID 1342213445.
The pointed observations were carried out both at low and high spectral resolutions (FWHM $\sim$ 30 GHz and 1.44 GHz),
covered with 32 repetitions each, for a total on-source integration time of 410 seconds and 4,262 seconds 
respectively\footnote[10]{See the Herschel SPIRE Observers manual, available at http://herschel.esac.esa.int/Docs/SPIRE/html/spire\_om.html}.
The shallower mapping observations at high resolution had 8 repetitions and covered an area
with a diameter of approximately 2$\arcmin$. The total on-source integration time was 17,050 seconds.

The FTS beam size and shape vary as a function of frequency and cannot
be characterized by a simple Gaussian response.
As can be seen from Figure~\ref{lines_fwhm}, where the observed source sizes 
of the brightest lines detected by the mapping observations of the SPIRE FTS 
are plotted as a function of frequency
(see the method described in Section~\ref{sec:sizes}), 
the FTS beam widths (FWHM) range from 17\arcsec~to 42\arcsec, probing a wider area in the sky with
increasing wavelength, and show a strong discontinuity between the
SLW and SSW bands, due to the multi-moded feed-horns used for the
spectrometer arrays (see Section 4.2.3 and Figure 5.13 in the SPIRE
Observers' Manual\footnotemark[10]).

The combination of the complex morphology of the inner region of
NGC\,1068 with the SPIRE-FTS beam properties yield a particular
coupling effect between the two.
Figure~\ref{CO_maps} shows the CO(3-2) and CO(1-0) interferometric maps from \citet{tsa12} of the 
central region of NGC\,1068.
As can be seen from these maps, the morphology of NGC\,1068 is dominated by two
main components \citep[see also, e.g.,][and references therein]{gar10}: the compact (of the order of 4\arcsec~in diameter) CND  
at the center of the galaxy, possibly associated with the active nucleus, 
and a relatively extended ring, with a radius of the order of 10-20\arcsec, whose emission
is dominated by star formation (SF) activity. 
The relative contribution from these two spatial components to the various rotational 
transitions of the $^{12}$CO molecule strongly varies with the transition 
from the low-J (J$_{up}\sim4$) to the high-J (J$_{up}\sim13$) lines. 
In this range the dominant source of CO excitation changes from the extended SF-ring to the compact CND. 
There is also a variation of the angular scale
probed by the FTS beam, from extended (beam FWHM $\sim$ 42\arcsec) to compact (17\arcsec),
as shown by the two concentric circles in the maps of Figure~\ref{CO_maps}.

\begin{figure}[ht!]
\includegraphics[width=\columnwidth]{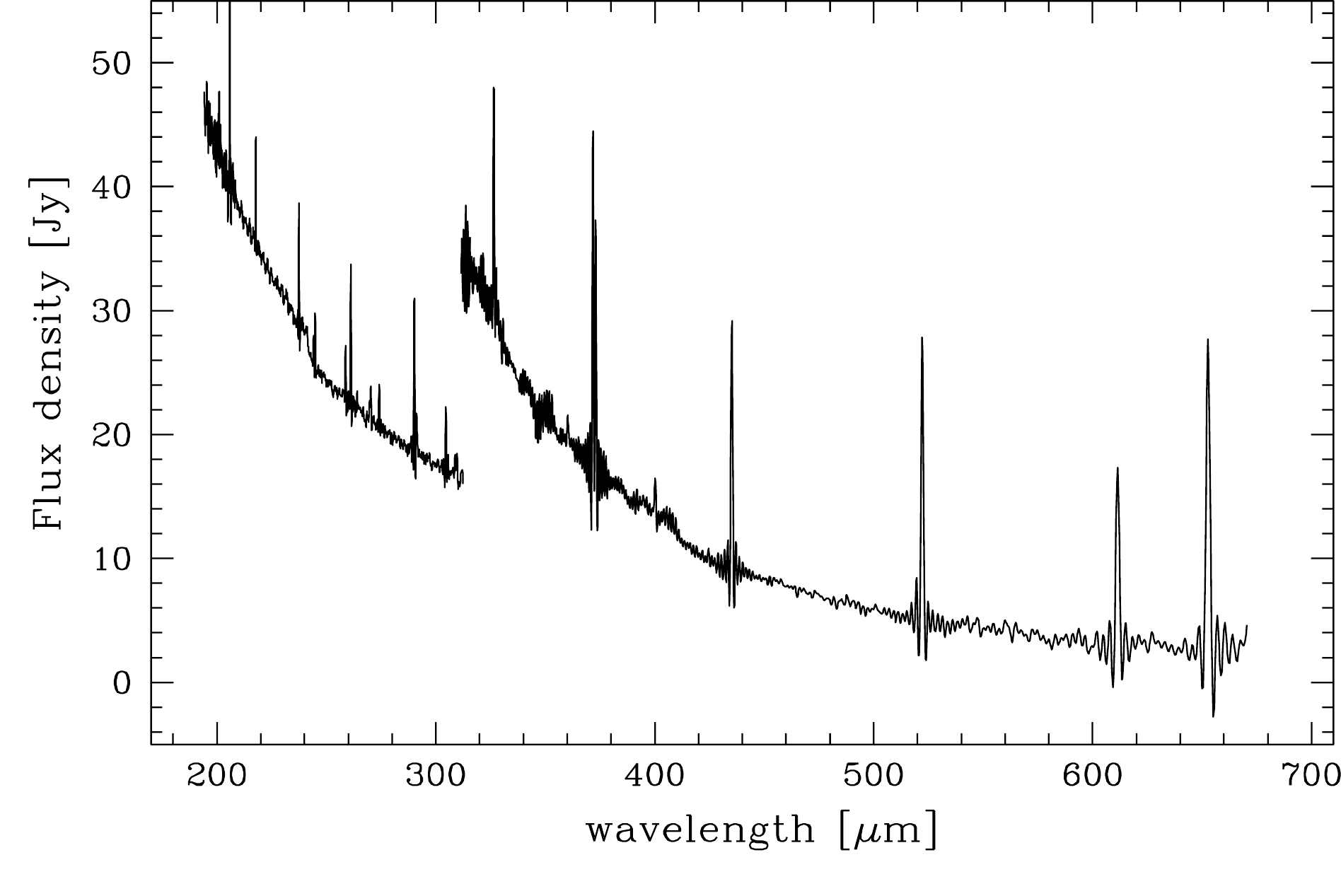}
\caption{The observed submillimeter spectrum of NGC\,1068, as measured from the SPIRE-FTS pointed observations. 
As can be seen the two SSW and SLW spectra do not overlap. This is due to both the %, because of the 
different beam sizes and the intrinsically different source sizes.}
\label{fig:n1068_cont}
\end{figure}

\begin{figure*}[ht!]  
\includegraphics[width=\textwidth]{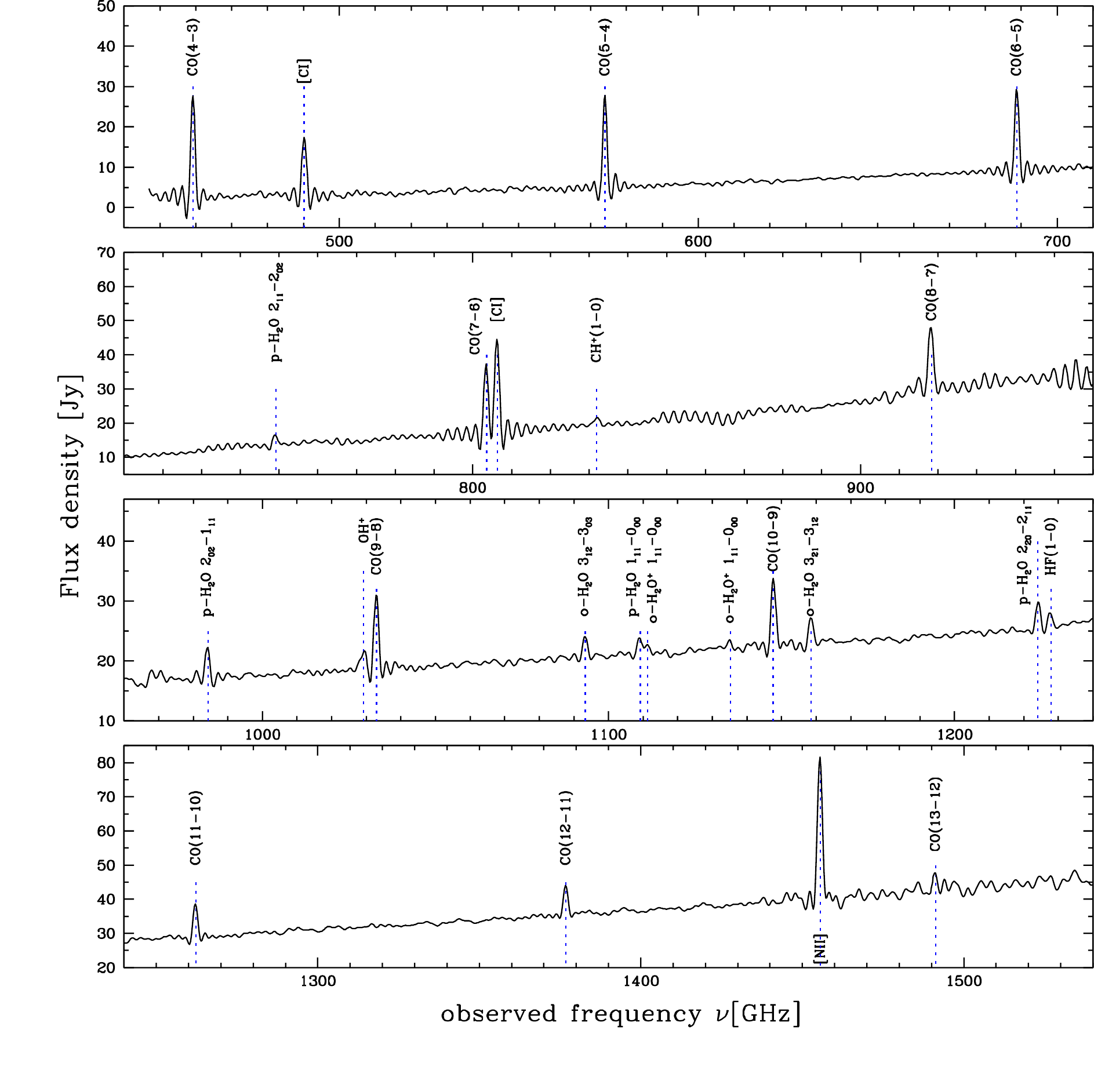}
 \caption{The observed submillimetre spectrum of NGC\,1068, as measured from the SPIRE-FTS pointed observations. From top to bottom are
  presented the consecutive spectral ranges. The emission line positions are indicated. 
   Line identifications are given for detections with S/N$>$3$\sigma$ (see Table~\ref{tab:lines}).}
  \label{fig:n1068_full}
\end{figure*}

\subsubsection{\label{sec:sizes}Source size measurements}

We first describe the reduction procedure that was used for source size measurements, while in Sections 
\ref{sec:maps} and \ref{sec:point} 
we describe the reduction procedure for spectral line measurements. We employ two reductions as the two measurements have different
requirements.

The mapping observations of NGC\,1068 have been reduced with the standard reprocessing pipeline in HIPE v.7 
and converted into spectral cubes using the \textsc{NaiveProjection} task\footnote{This and the  following routines 
are described in the SPIRE Data Reduction Guide, available at: 
http://herschel.esac.esa.int/hcss\-doc\-6.0/load/spire\_dum/html/spire\_dum.html} 
with custom pixel sizes of 11\arcsec~ and 13.8\arcsec~ for the SSW and SLW bands, respectively.
These sizes were chosen empirically starting from a large size and reducing it until holes 
(pixel with zero coverage) appeared in the maps.
The final cube for each of the two FTS bands has a variable on-source integration time in the map, between 16 
and 32 FTS scans.
From these two cubes a number of sub-cubes have been obtained by extracting a 60 GHz wide window 
(corresponding to $\sim 200$ bins in high spectral resolution)
around each line detected in the spectrum of the brightest spaxel, 
where a spaxel is defined as each one of the 5x5 spatial pixels which fill the spectrometer field of view\footnote{See the PACS Observer's Manual at: http://herschel.esac.esa.int/Docs/PACS/html/pacs\_om.html},
and modeling and subtracting the continuum emission with a 1$^{st}$ order polynomial using HIPE's task
\textsc{LineIntensityMap}.

The FTS data from a pointed observation have two available
calibrations which provide the target's correct flux for the two
extreme cases of an unresolved source and of an extended source that
completely and uniformly fills the instrument beam.
To determine which calibration to use, we measured the emitting region size for several lines from the analysis of the spectra cubes
(see Figure~\ref{lines_fwhm}).

The source sizes of all the transitions present in the SSW band, with the exception of the [NII]205~$\mu$m line, 
are consistent with originating from an unresolved source and they are 
therefore well calibrated in the pointed spectrum using the unresolved
source solution.
For this reason, we used the line fluxes measured from the pointed
spectrum for all the transitions in the SSW band except for the [NII]205~$\mu$m line,
for which we have used the mapping observations.
([NII] emission arises from ionized gas and so we may expect some
contribution from the SF-ring.)
For all the transitions in the SLW band, we measured the source sizes from the
mapping observations.

We have obtained line integrated maps with the task \textsc{IntegrateMapFromCube} from the spectral sub-cubes.
The integration over the lines has made no use of the fitting functionality inside the task, because of
the possible spurious detections observed for spectral features with low signal-to-noise ratios.
These line maps have been used to measure the angular extent of the region emitting every detected line using a 2D Gaussian fit.
While the instrument PSF is not strictly gaussian, because a large fraction of its light is
in wider secondary lobes, this approach is robust enough to
quantify the scale size of the emitting region through the fit of the PSF core. 

In conclusion, the deep pointed spectrum in the SSW band samples primarily, not exclusively, the lines emitted by the compact source,
while the SLW band measures lines emitted by both the compact and the extended structures.
This coupling effect between the beam and the target morphology makes it 
difficult to reconstruct the correct spectral line
energy distribution (SLED) for the $^{12}$CO molecular lines for the CND alone.
To overcome these problems, we have combined the information extracted from the
pointed spectrum with those derived from the mapping observations.

\subsubsection{\label{sec:maps}Spectroscopic mapping observations}

The mapping observations of NGC\,1068 were again reduced using a modified version of the standard 
Spectrometer Map pipeline script in the \emph{Herschel} Interactive Processing Environment (HIPE) v.9, 
developers build 588 and version 8.1 of the SPIRE calibration context.  The standard reduction assumes 
that the source is extended and uniformly fills the beam. Since our
source does not fill the beam, the unique point source 
flux conversion measured for each bolometers was applied to each of
the bolometers in our two arrays. After the point source 
correction, spectral cubes for the SSW and SLW were created using the \emph{spireProjection} task with 
\emph{projectionType="naive"} with a custom pixel size of $11''$ and $13.8''$ respectively as before. 
This cube was used for all SLW and the [NII]$205\mu$m line intensity measurements.

\subsubsection{\label{sec:point}Pointed observations}	

The single pointing/sparse observations of NGC\,1068 have been reduced using the Spectrometer Single Pointing user pipeline 
 in the \emph{Herschel} Interactive Processing Environment (HIPE) v.9, developers build 588 and version 8.1 
 of the SPIRE calibration context. An older version of this pipeline is described in \citet{ful10}. 
 The observed submillimeter spectrum of NGC\,1068, as measured from the SPIRE-FTS pointed observations, 
 is shown in Figure~\ref{fig:n1068_cont}, where the large offset at $\sim 300 \mathrm{\mu m}$ is due to the large 
 difference in the beam of the SSW and the SLW spectrometers.

\subsection{Line intensities}

As explained in the previous section, the transitions measured in the SSW band, except for the [NII]205~$\mu$m line, have been 
considered spatially unresolved and therefore measured from the pointed/deep
spectrum, while, on the contrary, the other partially extended lines have been measured from the mapping observations.

In order to measure the line fluxes of the extended component, the line maps have been convolved with  
2D kernels \citep{ben12} to reach the same beam size %PSF 
at every frequency. The final PSF corresponds to the coarsest angular resolution (FWHM = 42\arcsec) 
reached by the FTS for a $^{12}$CO line (i.e. the CO (4-3) at 461.04 GHz).
The kernels were built using the SPIRE-FTS PSF measurements, dividing
the PSF image of the CO (4-3) line by that of each detected line in the Fourier space. 

The baseline was removed by first masking all of the spectral lines before 
fitting a high order polynomial to the remaining continuum. 
The intensities of the detected lines have been computed with a Levenberg Marquardt fitting procedure
using the \textit{sinc function} model. 
Table~\ref{tab:lines} presents, for each detected transition, the rest frequency, the energy of the upper level (in K),  
the beam FWHM (in arcsec.)(see the SPIRE Observers'  Manual\footnotemark[10],
the intensity value with both the 1$\sigma$ statistical uncertainty and total uncertainty, which include the
calibration uncertainty added in quadrature. 
For the mapping observations, we have adopted a fixed FWHM of 43.4\arcsec~ for all detected lines. 
For the pointed observations of the compact source, we have adopted a calibration uncertainty of 10\%, while
for the mapping observations of the extended source, in addition to the calibration uncertainty, we have also 
added in quadrature a flat-fielding uncertainty of 7\%(see the SPIRE Observers'  Manual\footnotemark[10]). 
The lines detected with the FTS are not spectrally resolved. 

Figure \ref{fig:n1068_full} presents the observed spectrum of NGC\,1068 with the identifications of all lines detected at S/N$>$3. 
A total of 24 emission emission lines have been detected with the \her -SPIRE FTS: ten lines from CO (from J$_{\rm up}$=4 to 
J$_{\rm up}$=13), six lines from \water, two from \hdop, one from \ohp, HF and CH$^+$ each, in addition to the three atomic lines (from [NII] and 
[CI]).
The apparent ringing in the spectrum is due to the \textit{sinc} response of the FTS 
to each line.  

In Table~\ref{tab:lines} we also included 
literature measurements, relative to both the compact source and the extended source. In particular,
we considered in our analysis the interferometric observations of CO (1-0), (2-1) and (3-2) from \citet{kri11}, 
and the water lines from the \her -PACS observations (S. Hailey-Dunsheath 2012,  private communication)
for the compact component, 
and, for the extended component, CO (2-1) Z-Spec spectrometer observations at the Caltech Submillimeter
Observatory (CSO) \citep{kam11} and 
James Clerk Maxwell Telescope (JCMT) CO (3-2) observations convolved
to the 43.4\arcsec beam (C. Wilson 2012, private comm.).

\section{Results and discussion}

\subsection{CO Radiation transfer modeling}

To obtain a general solution for the CO physical conditions, we have first used the 
radiative transfer code RADEX \citep{vdt07}.
Then, considering the best possible ranges of the parameters as derived from RADEX, we have also used 
the PDR and XDR models from \citet{wol10} and \citet{mei05} and \citet{mei07}, respectively,  
to provide constraints on the origin of the excitation of the measured lines, and in particular to determine 
whether there is an influence of the AGN, through its X-ray emission, on the molecular gas in the CND and SF-ring associated
with NGC\,1068.

\subsubsection{RADEX models \label{radex}}

RADEX is a non-Local Thermodynamic Equilibrium (LTE) code  
available from the Leiden Atomic and Molecular Database \citep[LAMDA, ][]{sch05}. 
Under the assumption of a uniform medium, 
using an escape probability formalism that models the entire emitting region as a single zone, 
RADEX performs statistical equilibrium calculations involving collisional and radiative processes.
We used the collisional rate coefficients with H$_2$ of \citet{yan10}. 
Considering the radiation field from background sources given as input, it computes the molecular level
populations in the optically thin limit in the first iteration 
and the optical depths and the escape probabilities for each line.
The interdependence of the molecular level populations and the local radiation field requires that
the solution is reached through an iterative method. At each iteration, the
code computes new level populations using new optical depth values, until the two converge
on a consistent solution. Then the resulting line intensities are given as output.

The input parameters required for RADEX are the background radiation field,
the kinetic temperature T$_{kin}$, the molecular hydrogen number density n(H$_2$)
(assumed to be the only collision partner),
and the ratio between the column density of the molecule and the width of the lines, 
which is the physically relevant quantity that determines the optical depth. 

We used RADEX to create a grid of solutions for the CO transitions spanning wide ranges in 
T$_{kin}$, n(H$_2$), and N$_{CO}$ (see Table \ref{table:radex}). % XXX replaced N_mol with N_CO
To compare this grid of models with our CO measurements we used the Bayesian likelihood analysis 
code \citep{war03}, developed by the Z-Spec Team 
\citep{nay10b,kam11}.

 \begin{figure}[hb!]
  \includegraphics[width=\columnwidth]{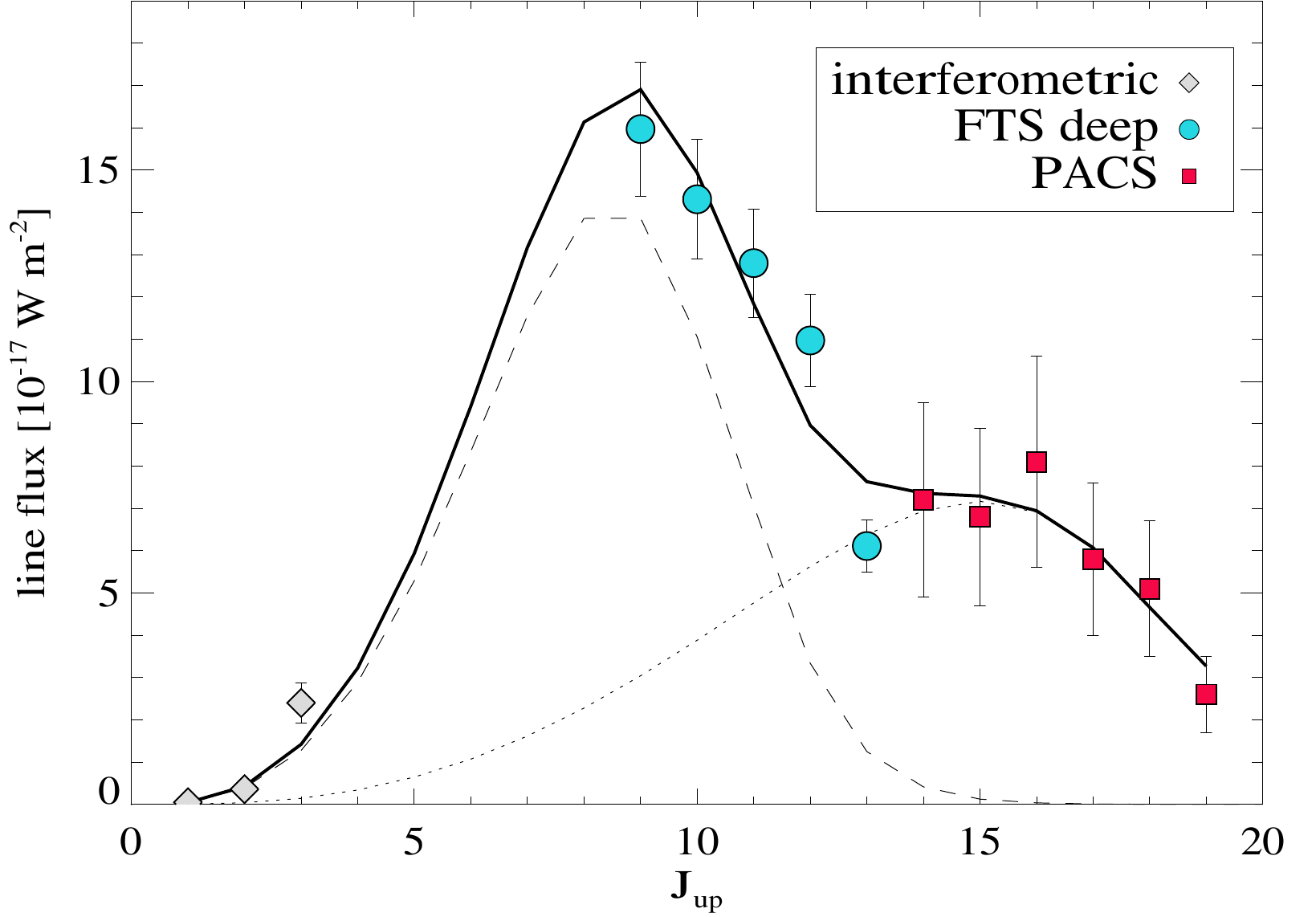}
 \caption{RADEX model fitting for the compact component associated with the CND.
Square points (red) represent the PACS data, circle points (blue) depict the FTS deep spectrum data, 
 and the diamond points (grey) show the interferometric observations of the CND \citep{kri11}, 
 the dotted line shows the ME model of \citet{hai12}, while
the dashed line shows the best fitting configuration of RADEX parameters 
(see column 4D Max of Table \ref{table:radexresults}). The sum of the two models is shown as
a solid line.
}
\label{compact_radex}
 \end{figure}

  \begin{figure}[ht!]
 \includegraphics[width=\columnwidth]{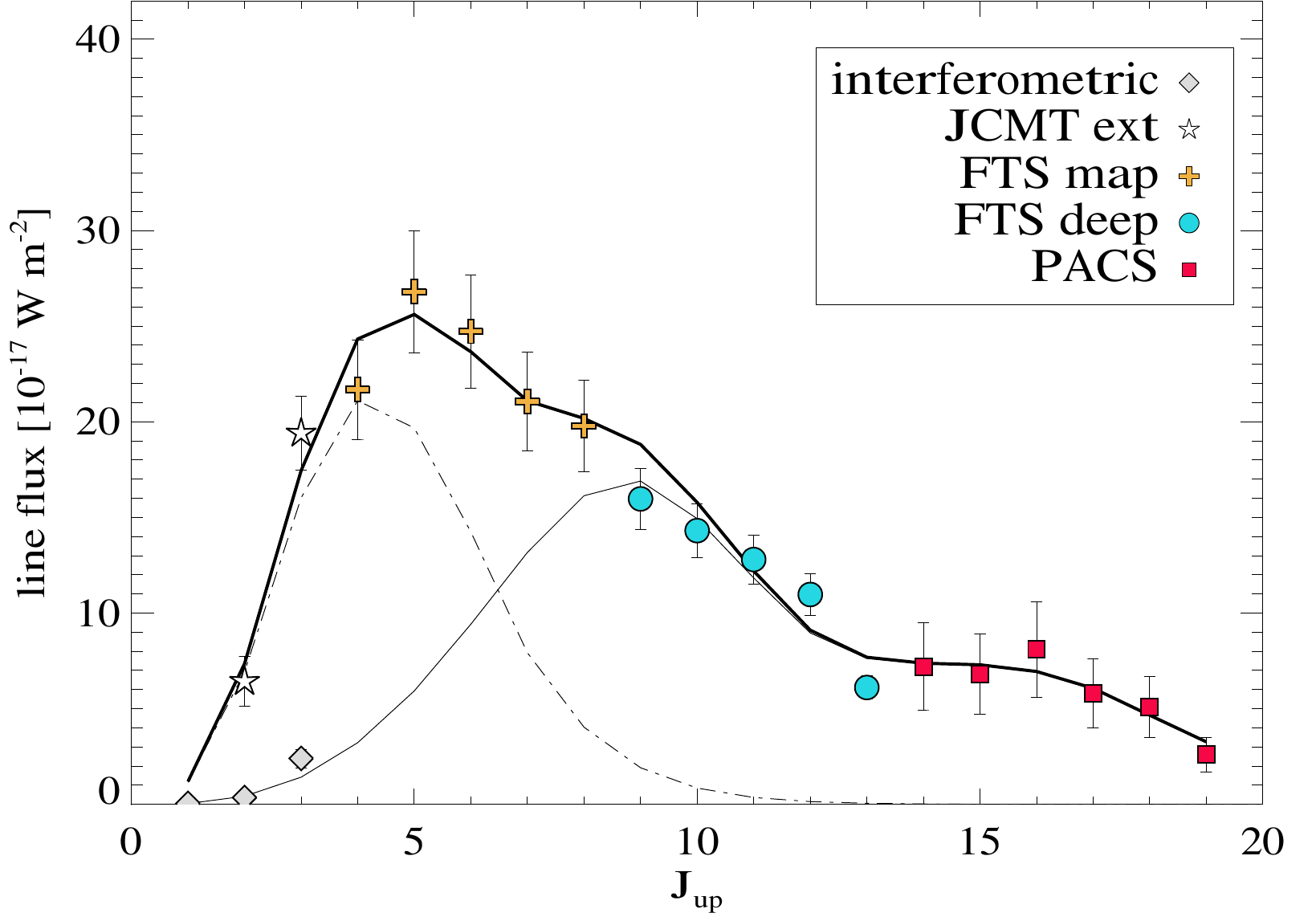}
 \caption{RADEX model fitting for the extended component associated with the SF-ring.
Notations as in Figure~\ref{compact_radex}. 
The crosses (yellow) represent the FTS measurements from the spectral map and the 
white stars show the ground-based observations of the extended component 
(\citealt{kam11},  Wilson 2012, private communication).
The dot-dashed line shows the best fitting configuration of RADEX parameters 
(see column 4D Max of Table \ref{table:radexresults}). The thin solid line is the best model for the 
CND component. The sum of all models is shown as a thick solid line.}
\label{extended_radex}
 \end{figure}
 
Computing the RADEX models, we have considered both models with and without 
the inclusion of the local background radiation to be added to the cosmological component of the CMB at 2.73K.
To estimate the local background in the far-IR, we have fitted the aperture photometry of the \her -PACS 
and SPIRE maps at 70, 160, 250, 350 and 500~$\mu$m, respectively, 
that will be presented and discussed in a forthcoming paper (Spinoglio et al. 2012, in prep.). We have  
used a circular aperture with a diameter of 33$\arcsec$~ and a dust emissivity law with $\beta$=2
and we have derived a gray-body temperature of 35 K.
In agreement with the higher J-CO line fitting results of \citet{hai12}, we have found that the inclusion of this local background 
does not change significantly the results of the CO fits
that only include the CMB. Therefore we have adopted only the CMB black-body radiation at 2.73 K
as background radiation, for both the extended and compact sources. 

The presence of several spatial and spectral components in the CO SLED has required a multi-step approach,
looking for the best solution of one component at a time and subtracting it from the others.
We started by considering the CO SLED of the compact source, composed of the FTS-SSW 
measurements in combination with the compact interferometric observations. 
For the high-J (J$_{\rm up} \ge 14$) PACS data, we used the medium excitation (ME) component LVG model 
presented by \citet{hai12}, characterized by a density of $10^{5.7}\
{\rm cm^{-3}}$ and a kinetic temperature $T_{kin}=150$~K.
The ME model for the J$_{\rm up}$ $>$ 13 CO lines was subtracted from the observed CND
spectrum before attempting to fit the SLED. 

In Table \ref{table:radexresults}, the results obtained with the Z-Spec code are summarized.
For each of the two models of the CND and SF-ring the following parameters are given: the kinetic temperature of the CO,
$T_{\rm kin}$, the molecular hydrogen density, $n({\rm H_2})$, the CO column density, $N_{CO}$, the area 
filling factor, $\Phi_{\rm A}$, the gas pressure, $P$, the beam averaged CO column density, $\langle N_{\rm CO} \rangle$,  and 
the molecular hydrogen mass M(H$_2$), that has been computed using a CO abundance of $x_{CO}= 3\times 10^{-4}$,
adopted by the Z-Spec code
\citep{nay10b,kam11}.

For each parameter, the median value, the 1-$\sigma$ interval, the 1D and the 4D maximum values are given.
1D refers to the maximum value of the integrated parameter distribution while 4D refers to the 
value of that parameter at the best fit solution.

The compact source is unresolved with all the considered instruments and identified with the CND, while the extended
source data probe the emission arising from a region of 43.4\arcsec~  in diameter, compatible with the SF-ring structure.
We have obtained the best fit to the CND with RADEX,  as can be seen in Figure \ref{compact_radex},
by fitting the mid-J lines (from J$_{up}$ = 9 to J$_{up}$  = 13) together with the low-J lines observed with interferometric
techniques, which isolate the CND component, and subtracting the contribution from the medium excitation (ME) component of the LVG model of \citet{hai12}.

The best fit of the SF-ring with RADEX (see Figure \ref{extended_radex})
was then obtained by subtracting from the observed lines (from J$_{up}$ = 4 to J$_{up}$  = 8
with SPIRE and J$_{up}$ = 3 and 2 with JCMT) the contribution arising from the compact source by 
summing the mid-J and high-J contributions. 
As can be seen in Figure \ref{extended_radex}, the extended source gives a significant 
contribution to the over-all SLED only for J$_{up}<$9. 
To summarize, as can be seen in Table~\ref{table:radexresults}, we found that the best fit RADEX model results for the CND give  $T_{\rm kin}$=100~K and $n({\rm H_2})$= $10^{4.5}$
cm$^{-3}$ and for the SF-ring   $T_{\rm kin}$=127~K and $n({\rm H_2})$= $10^{2.9}$
cm$^{-3}$.

\subsubsection{PDR and XDR models of the CND component \label{pdr_xdr_cnd}}

PDR and XDR models from \citet{wol10} and 
from \citet{mei05} and \citet{mei07}\footnote{Available at: http://www.strw.leidenuniv.nl/meijerink/grid/} 
describe the thermal and chemical balance of molecular gas that is exposed to 
far-ultraviolet (FUV) radiation (6-13.6 eV) and X-rays (1-100 keV), respectively.
The codes used to determine the gas conditions in these regions as a function of depth 
take into consideration elaborate chemical networks and the cooling, heating and chemical processes
which are completely determined by the radiation field.

\begin{figure*}[ht!]
  \includegraphics[width=\columnwidth]{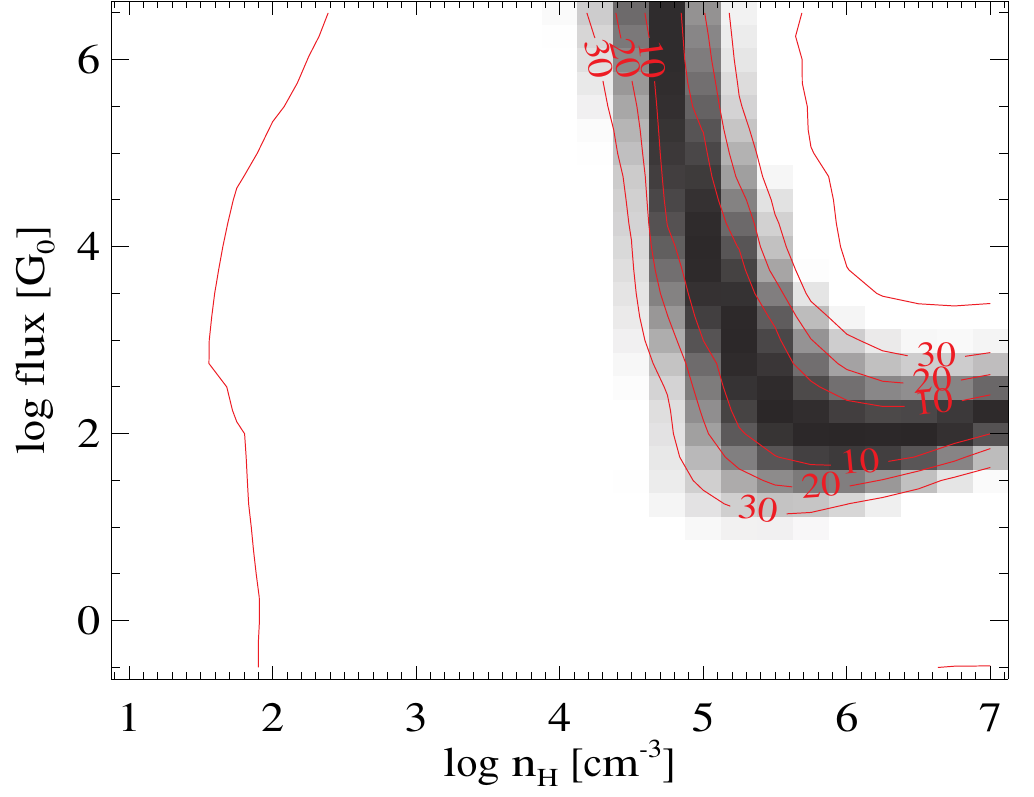}
  \includegraphics[width=\columnwidth]{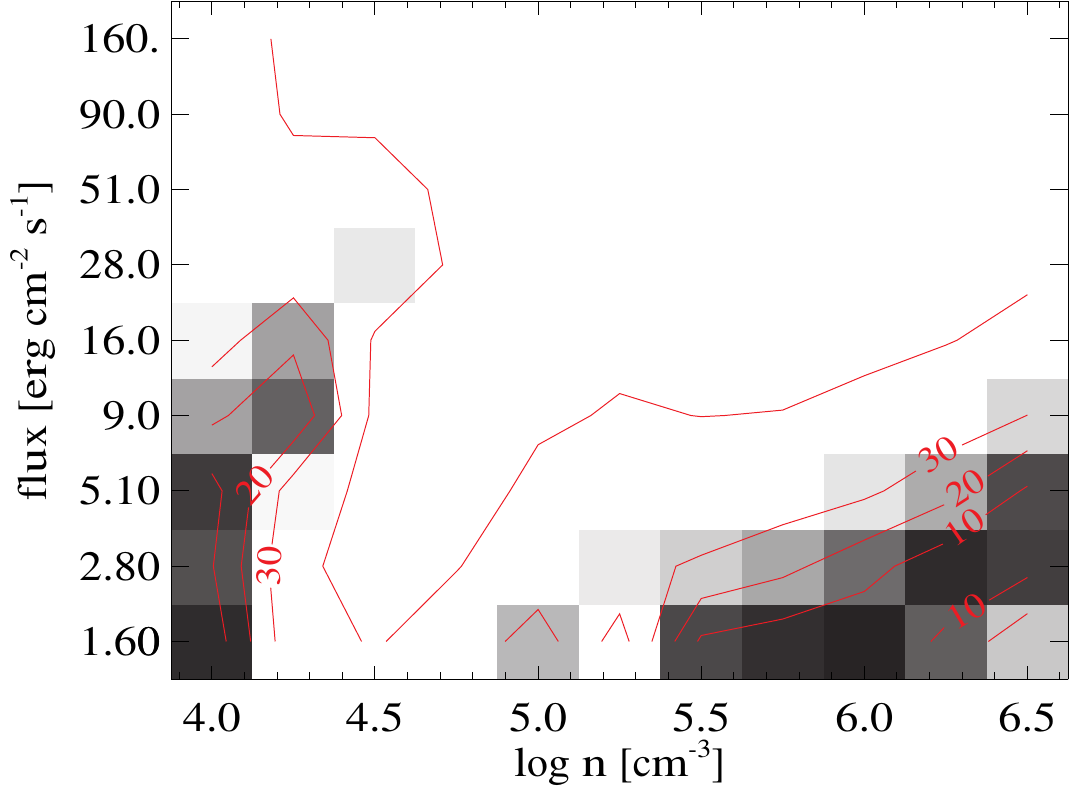}
 \caption{Reduced $\chi^2$ values of the  models for the CND.
  {\it Left:} PDR models \citep{wol10} ; {\it Right:} XDR models \citep{mei05,mei07}. }
\label{chi_cnd}
\end{figure*}

We have considered the grids of PDR models of \citet{wol10},  
that span from 10$^1$ to 10$^7$ cm$^{-3}$ in density and from 
10$^{-0.5}$ to 10$^{6.5}$ G$_0$\footnote{G$_0$ = 1.6$\times$10$^{-3}$ 
erg cm$^{-2}$ s$^{-1}$ and the field is integrated from 6 to 13.6 eV.} in FUV flux.
The PDR models are based on those of \citet{kau06}, 
with updates from \citet{wol10} and \citet{hol12}.  
In particular  the atomic and molecular freeze-out and grain chemistry in \citet{hol12} are included,  
as well as their PAH photo rates which affect the ion-neutral chemistry and the production of CO. We use their
``linear yield" PAH$^0$ photoionization rate slightly modified for a \citet{dra78} 
interstellar radiation field. Hot CO production by HCO$^+$ recombinations is also included, as suggested by J. Black (2012, private 
communication).

For the XDR models,  we used the grid by \citet{mei07} that spans two different ranges in density
($10^3 - 10^{4}\ {\rm cm^{-3}}$ and $10^4 - 10^{6.5}\ {\rm cm^{-3}}$) irradiation 
($F_{X} = 1.6 - 160\ {\rm erg s^{-1} cm^{-2}}$), and cloud size of 1 pc and 10 pc. 
We used a $\chi^2$ minimization algorithm to identify the best description of the $^{12}$CO line fluxes
measured with SPIRE and interferometric data (see Table ~\ref{tab:lines}).

As already mentioned in Section ~\ref{radex}, in addition to the CO lines detected in NGC\,1068 by the SPIRE FTS, presented in Table~\ref{tab:lines},
higher J lines (from J$_{up}$ = 14 to J$_{up}$  = 24) have been observed by PACS and presented in \citet{hai12}. 
We have tried to model the whole sequence of CO lines detected with PDR/XDR models, however no single component
was able to fit the data;  three distinct components were needed, all originating from the compact CND, 
to fit the CO lines down to J$_{up}$=9. 
Moreover, as outlined in \citet{hai12}, even if - in principle - PDR models could fit the data from J$_{up}$ = 14 to J$_{up}$  = 24, 
the morphology of the H$_2$ near-IR emission and the two different kinematics of the CO lines 
at J$_{up}\leq$ 17 and J$_{up}\geq$ 20, indicate that these two sets of lines 
trace physically distinct components, therefore excluding a PDR origin of these high-J CO lines.
Consequently, for the high-J (J$_{up} \geq 14$) PACS data we used the model presented by \citet{hai12}, consisting of
two XDR components with a density of $10^{5.75}$\,cm$^{-3}$ and incident X-ray fluxes of 9 and 160 erg\,s$^{-1}$\,cm$^{-2}$, respectively.
Then we have performed the analysis %was executed 
on both the compact source and the extended source, once the other components were
fitted and removed as explained in the previous section.

The mid-J lines originating from the CND are best described by a sequence of PDR models ranging 
from an incident FUV flux of G$_{o}$ $\sim$ $10^{5.0}$ and density of $10^{4.5}$\,cm$^{-3}$, 
in agreement with the RADEX model, to a FUV flux  of G$_{o}$ $\sim$ $10^{2.0}$
and density of $\sim$ $10^{6.0}$\,cm$^{-3}$ (see left panel of Figure \ref{chi_cnd}).
On the other hand, also XDR models either with density $\sim 10^{4.0}$\,cm$^{-3}$ or $\sim 10^{6.0}$\,cm$^{-3}$ and incident X-ray flux in the range 
of 1.6--5.0\,erg\,s$^{-1}$\,cm$^{-2}$ 
are able to reproduce the observed mid-J CO fluxes with a good $\chi^2$ value (see right panel of Figure \ref{chi_cnd}). 
However, the RADEX models (see section ~\ref{radex}) %exclude any solution with a density as high as $10^{6.0}$\,cm$^{-3}$. 
favor solutions with n$_H$ between $10^{4.4}$ and $10^{4.8}$\,cm$^{-3}$ as listed in Table \ref{table:radexresults}. %3

\begin{figure}[h!]
  \includegraphics[width=\columnwidth]{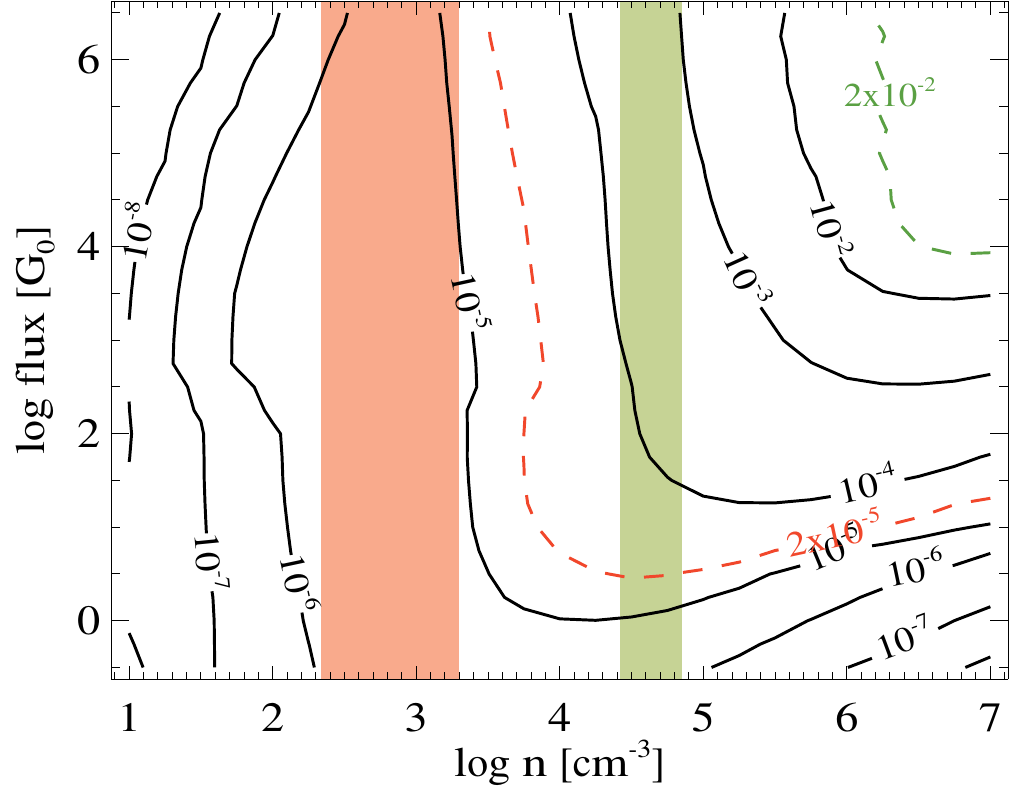}
 \caption{Total CO surface brightness (in units of erg~s$^{-1}$~cm$^{-2}$~sr$^{-1}$) 
 as a function of incident FUV flux and density for the PDR models. 
 The left shaded region (in orange) shows the 1$\sigma$
 range of the density of the best fit RADEX model of the SF-ring (extended region),  while the right shaded 
 region (in green) reports the area of the 1$\sigma$ range of the density of the best fit RADEX model 
 of the CND (compact component). The red dashed line represents the measured surface density 
 for the SF-ring, while the green dashed line represents the measured surface density 
 for the CND of 0.02~erg~s$^{-1}$~cm$^{-2}$~sr$^{-1}$.  }
\label{surface_pdr}
\end{figure}

\begin{figure}[h!]
  \includegraphics[width=\columnwidth]{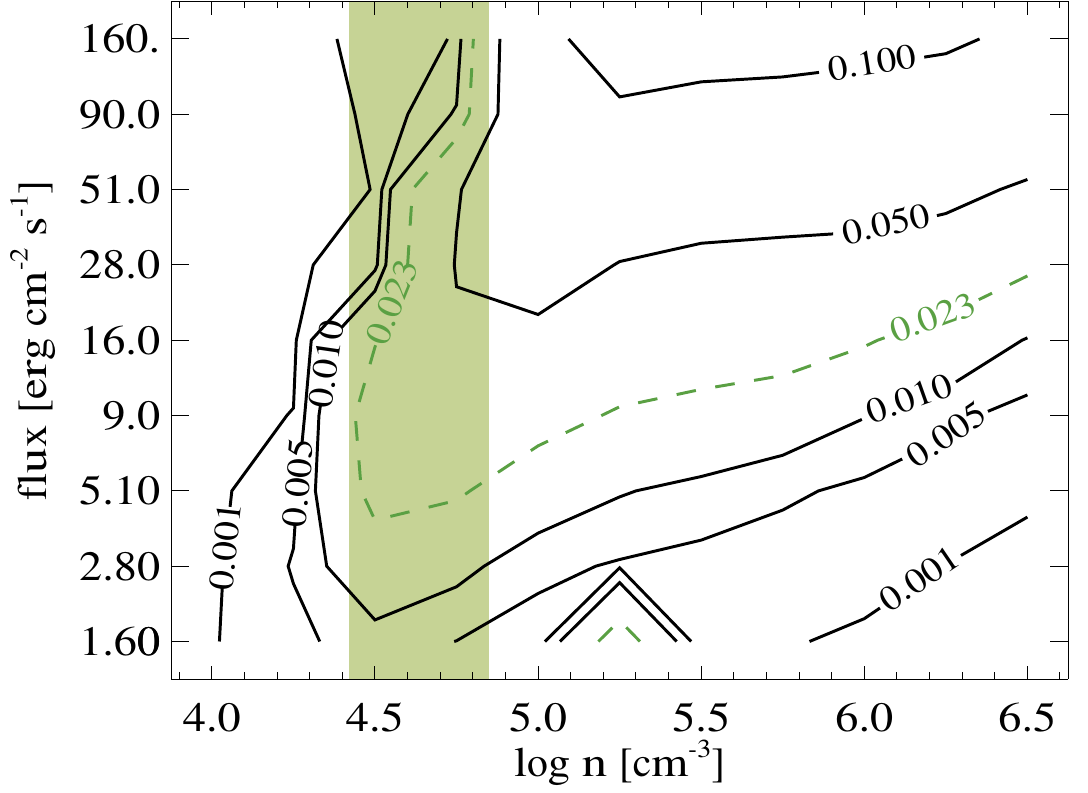}
 \caption{Total CO surface brightness (in units of erg~s$^{-1}$~cm$^{-2}$~sr$^{-1}$) 
 as a function of incident FUV flux and density for the XDR models. 
 The shaded region (in green) shows the 1$\sigma$ range of the density of the best fit RADEX model 
 of the CND (compact component). The dashed line (in green) represents the measured surface density.}
\label{surface_xdr}
\end{figure}

\begin{figure}[ht!]
  \includegraphics[width=\columnwidth]{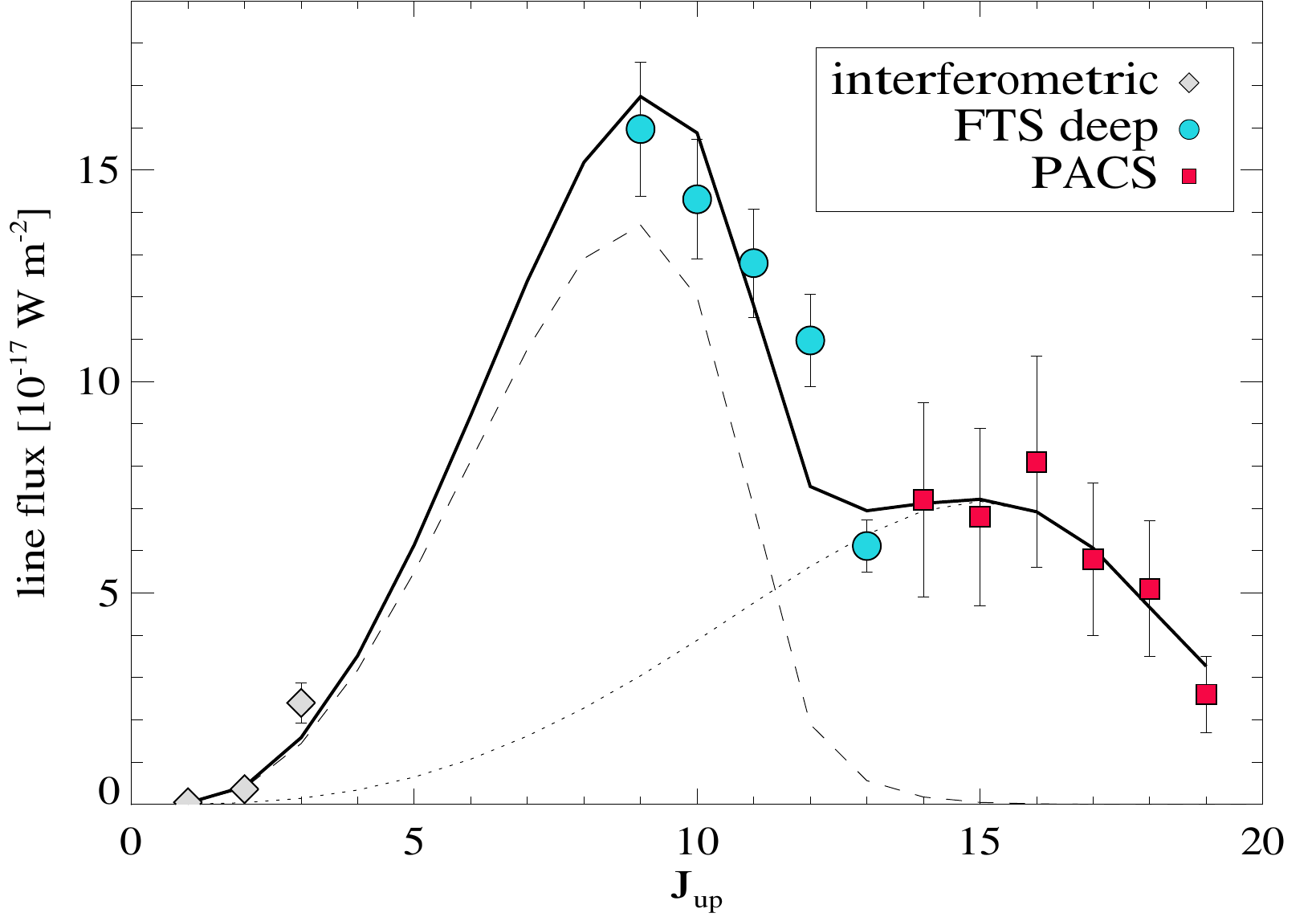}
 \caption{XDR model fitting for the compact component, associated with the CND. 
 Models have been taken from \citet{mei05} and \citet{mei07}. 
 The points follow the same color code as in Figure \ref{compact_radex}.
 The dotted line shows the fit of the high-J CO 
 lines with the XDR model of \citet{hai12}, the dashed lines show the fit of the intermediate-J CO lines, 
 with a XDR model with density $n=10^{4.25} cm^{-3}$ and X-ray flux of 9~erg~s$^{-1}$~cm$^{-2}$, 
 once the high-J CO XDR model has been subtracted. 
 The solid line shows the total modeled emission, which is compared to the observed data (see the text).}
\label{compact_PDR}
 \end{figure}

To further constrain the models, we have used the CO surface brightness for both PDR 
and XDR models, by summing over all of the modeled CO lines. 
For the CND, we have computed the total CO emission from the best fit XDR model 
presented in Figure ~\ref{compact_PDR} and divided by the emitting area of the CND, assumed to be 4~$\arcsec$ in diameter, 
and corrected by the filling factor of the RADEX best fit (see Table~\ref{table:radexresults}) to derive the CO surface brightness of the CND.

In Figures ~\ref{surface_pdr} and ~\ref{surface_xdr} we present the total CO surface brightness for both PDR and XDR models, respectively,
with the density region allowed by the RADEX models indicated by shaded regions.
From Figure ~\ref{surface_pdr}, the predicted CO surface brightness in the CND by PDR models, which satisfies the density 
constrain of the RADEX model, 
is about two orders of magnitude below the value computed with the best fit model of the CND 
of 0.02~erg~s$^{-1}$~cm$^{-2}$~sr$^{-1}$. 
On the other hand, Figure ~\ref{surface_xdr} shows that the predicted CO surface brightness   
in the CND, for an XDR model with density of $\sim$ 10$^{4.3}$ cm$^{-3}$ and incident X-ray flux 
$\geq$ 3 erg~s$^{-1}$~cm$^{-2}$,
agrees well with the best fit value of 0.02~erg~s$^{-1}$~cm$^{-2}$~sr$^{-1}$. 

We have also compared the observed parameters of the CND in NGC\,1068 with the XDR model results presented
for Arp~220 in \citet{ran11}. The XDR model of Arp~220 has been computed
using an updated version of the code described by \citet{mal96}. The assumed physical conditions in Arp~220 coincide
with those of NGC\,1068: a hard X-ray (1-100keV) luminosity of L$_{X}$ $\sim$10$^{44}$erg~s$^{-1}$, a power law index 
of $\alpha$ =0.7  and an absorbing hydrogen column density of N$_{H}$ = 10$^{24}$~cm$^{-2}$, 
typical of Compton thick sources, like NGC\,1068. 
In their Figure 7 (bottom, right panel), \citet{ran11} show the predicted CO surface brightness
as a function of density and radius of the emitting region. Our best fit  value of 0.02~erg~s$^{-1}$~cm$^{-2}$~sr$^{-1}$
is in reasonable agreement with a density of $\sim$10$^4$~cm$^{-3}$ and a radius of 100-200~pc. The radius that we associate to
the CO emitting region in the CND is about 160~pc. We therefore conclude that the XDR origin of the CO CND emission in 
NGC\,1068 is indeed a plausible explanation.
In Figure ~\ref{compact_PDR} we show the fit to the observed CO fluxes using the best fit XDR 
model with the adopted density and X-ray flux for the CND component.

\subsubsection{PDR and XDR models of the SF-ring component}

\begin{figure*}[ht!]
  \includegraphics[width=\columnwidth]{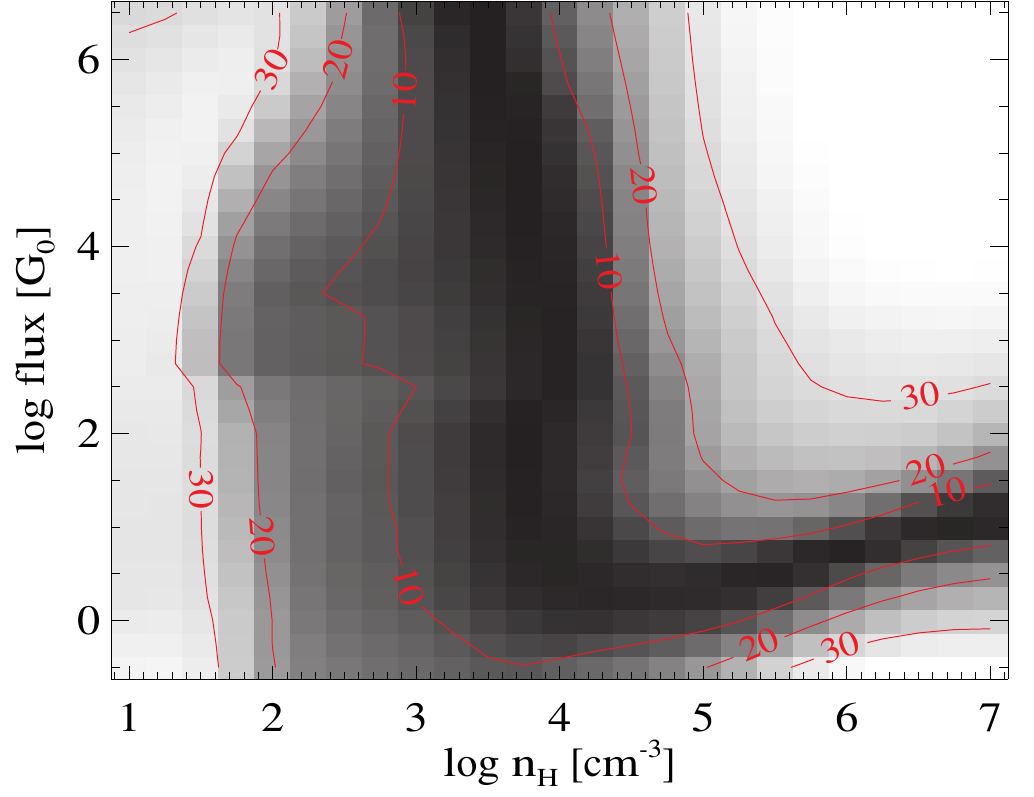}
   \includegraphics[width=\columnwidth]{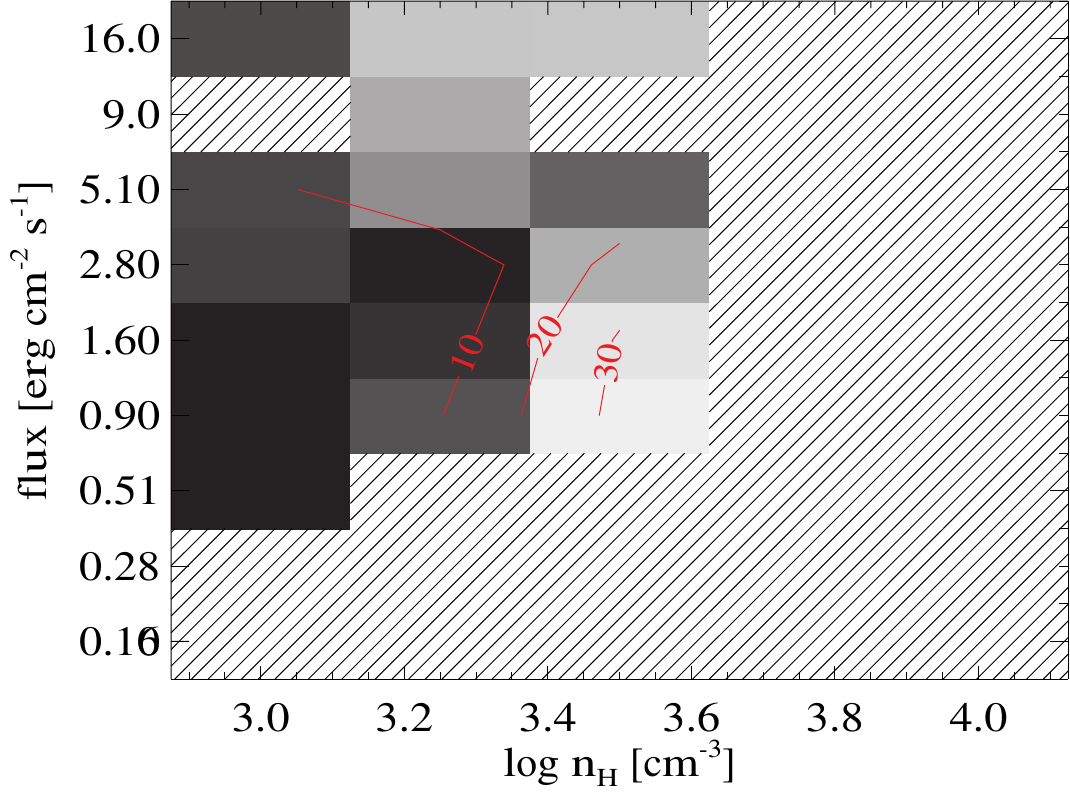}
 \caption{Reduced $\chi^2$ values of the PDR \citep{wol10} and XDR models \citep{mei05,mei07} for the extended component.
 {\it Left:} PDR models; {\it Right:} XDR models. The line shaded 
region of the figure is where no model is available or the model does not predict the lines of interest.}
\label{chi_wide}
\end{figure*}

The low-J component of the SLED can be fitted either by PDR models with a FUV flux  of G$_{o}$ $\gtrsim$ $10^{2}$
and density of 
$\sim$ $10^{3.5}$\,cm$^{-3}$, in rough agreement with the RADEX modeling results, or with 
models with lower FUV flux (G$_{o}$ $\sim$ $10^{1}$) and a higher density of $\gtrsim$10$^{4}$\,cm$^{-3}$,
which are, however, in conflict with the RADEX results. 
Moreover (see the right of Figure \ref{chi_wide}), XDR models with a
 density $\sim 10^{3.0}$\,cm$^{-3}$ and incident X-ray flux in the range of 0.5--2.0\,erg\,s$^{-1}$\,cm$^{-2}$ 
can fit the data.
 
  \begin{figure}[ht!]
 \includegraphics[width=\columnwidth]{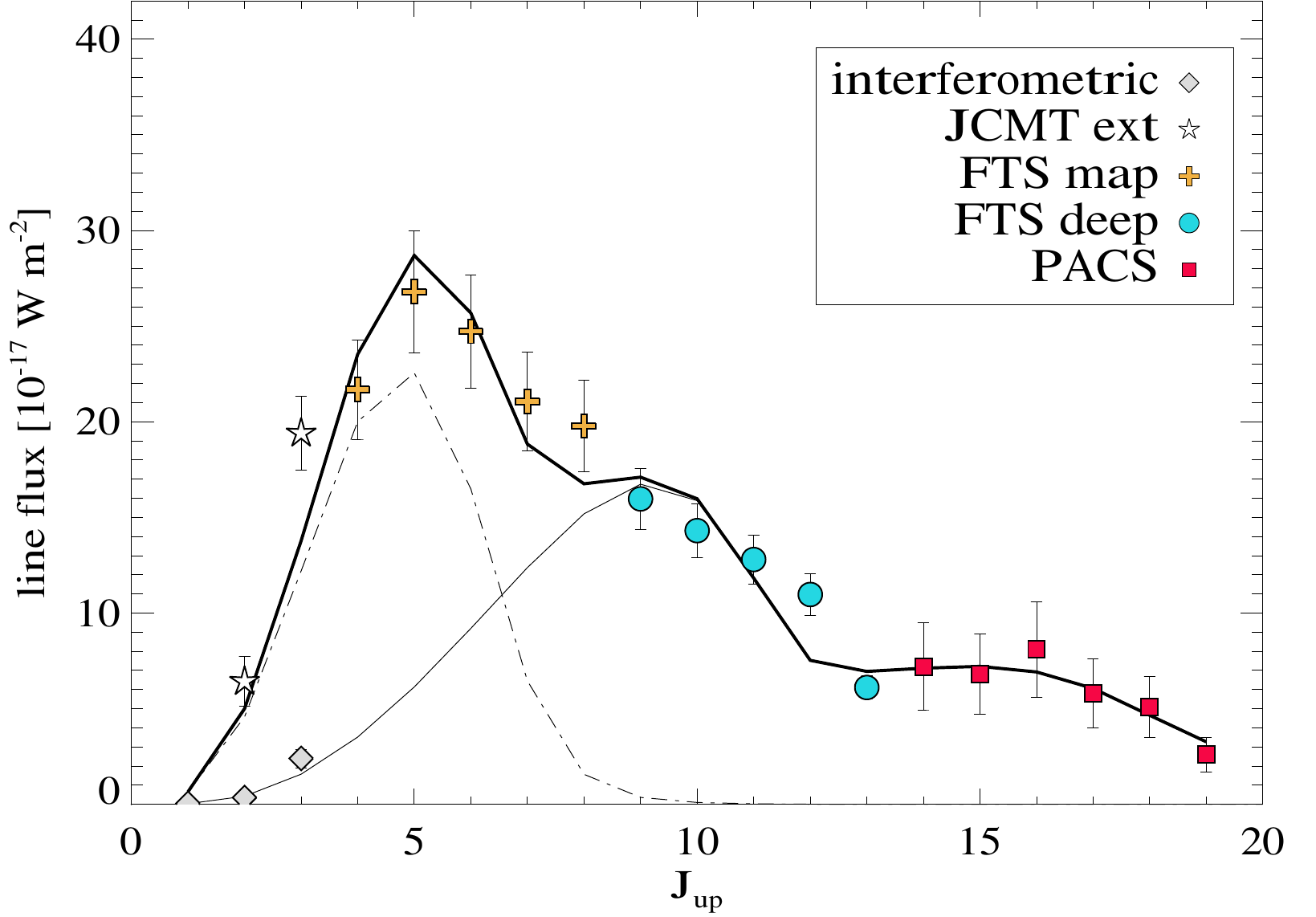}
 \caption{PDR/XDR model fitting for the extended component (SF-ring), 
 together with the compact component, associated with the CND.  
 Models have been taken from \citet{mei05,mei07}. 
 Notations as in Figure~\ref{extended_radex}. The crosses (yellow) represent the FTS measurements 
 from the spectral map and the stars (white) show the ground-based observations of the extended component (\citealt{kam11},  Wilson 2012, 
 private communication). The thin solid lines show the fit of the  compact component, which is the sum 
 of the high- and mid-J models as in Figure \ref{compact_PDR}. The dash-dotted lines show the fit of the extended component, 
 with a PDR model with density $n(H_2)=10^{3.25} cm^{-3}$ and log $Go=2.75$, 
 once the compact component has been subtracted. 
 The thick solid line shows the total modeled emission, which is compared to the observed data (see the text).}
\label{extended_PDR}
 \end{figure}

Similarly to the case of the CND, described in Section ~\ref{pdr_xdr_cnd},
for the SF-ring we have used the total CO emission from the best fit PDR model 
and the area of 40$\arcsec$ in diameter 
to derive the extended component surface brightness. For this latter case, we did not use any filling factor because we consider that 
the extended emission in the SF-ring is filling most of the beam.

To further constrain the models, we use again the CO surface brightness. 
From Figure ~\ref{surface_pdr},  the predicted CO surface brightness   
in the SF-ring, for a PDR model with density of $\sim$ 10$^3$ cm$^{-3}$ and incident FUV flux above $G_0\geq10^2$,
agrees well with the observed value of $2\times$10$^{-5}$ erg s$^{-1}$ cm$^{-2}$ sr$^{-1}$,
reinforcing this solution. 
We present the best fit model in Figure~\ref{extended_PDR}. 

\subsubsection{Constraining the PDR and XDR models with luminosities}\label{pdr_xdr}

In order to verify that the XDR models used to fit the compact (CND) and extended (SF-ring) molecular 
components in NGC\,1068 are in agreement with the intrinsic X-ray flux emitted from the AGN,
we have adopted an estimated intrinsic luminosity in the range of L$_{1-100keV}$ $\sim$ 10$^{43}$-10$^{43.5}$ erg~s$^{-1}$
\citep[see, e.g.][]{col02}.
Using the diameter of the CND of 4$\arcsec$ and that 
of the SF-ring of 40$\arcsec$, we derived an X-ray flux in the range of $\sim$ 3-10 erg~s$^{-1}$cm$^{-2}$ and of  
$\sim$ 0.03-0.11 erg~s$^{-1}$cm$^{-2}$, respectively for the compact and extended components. 

The X-ray flux predicted for the XDR model of the CND component is  in the range 1.6$<$F$_{X}$$<$10 erg~s$^{-1}$cm$^{-2}$ 
(see right panel of Figure \ref{chi_cnd}), therefore the hypothesis of an XDR origin of the intermediate-J CO lines in the CND can be considered 
valid.
The X-ray flux predicted for the XDR model of the SF-ring of 
F$_{X}$ = 0.5--2.0\,erg\,s$^{-1}$\,cm$^{-2}$ is not %only in marginal 
in agreement with our estimates, being at least one order of magnitude higher, 
and therefore the XDR origin of the lower-J CO lines in the extended region can be excluded, 
given the assumption in the above paragraph.
Moreover, even if the X-ray  flux would be available to excite the XDR, 
e.g. in the case of an underestimated intrinsic X-ray luminosity, 
an unknown part of it is expected to be absorbed from %the torus 
the interstellar medium along the path, before reaching the SF-ring. 
The attenuation to the X-ray source as seen by the CO-emitting gas cannot in fact be quantified. 
Therefore the estimated flux could be considered as an upper limit.

Similarly we estimated the far-UV (6--13.6\,eV) flux from the AGN that illuminates the CND and the SF-ring. 
The adopted intrinsic far-UV luminosity of the AGN is $\sim$10$^{42}$\,erg~s$^{-1}$. This value was calculated 
using the intrinsic far-UV continuum derived by \citet{pie94}. At the distance of the SF-ring the far-UV flux is 
$F_{\rm FUV} = 10^{0.5}$\,G$_0$, whereas for the CND the $F_{\rm FUV} = 10^{2.4}$\,G$_0$.

This $F_{\rm FUV}$ for the CND implies a high-density ($n_{\rm H}> 10^{5.5}$\,cm$^{-3}$) component 
according to the PDR models (left panel of Figure \ref{chi_cnd}), which is not in 
agreement with the density estimate from the RADEX modeling.
For the extended component,  the UV flux from the AGN is lower than that predicted by the PDR models;  however,  
in the star-forming ring it is likely that the UV emission from young stars also contributes to the interstellar UV radiation field. 
For comparison, the far-UV flux from a young stellar cluster (age 5\,Myr, see \citealt{spi05}) of stellar mass 
10$^4$\,M$_\odot$ is 10$^2$\,G$_0$ at 50\,pc. This far-UV flux is an order of magnitude higher than that 
from the AGN and would be compatible with the predictions of the PDR models (see left panel of Figure \ref{chi_wide}).

Taking into account all the above considerations, we therefore conclude that the most plausible explanation for the excitation of the CND
is an XDR illuminated from the AGN, while that of the SF-ring is a PDR, mainly excited from the young stellar
populations in the galactic arms.

\subsection{H$_2$O emission}

We detected 6 \water\ transitions, 2 o-\water\ and 4 p-\water\ lines, above a 3~$\sigma$ level 
within the SPIRE--FTS spectrum. In Table~\ref{tab:lines} we list the detected lines, reporting their frequency, upper level energy 
$E_{\mathrm{up}}/\kappa_{B}$, 
and the measured flux. 
We also listed the water lines detected with PACS (S. Hailey-Dunsheath 2012, private communication). 
The line fluxes detected by PACS were measured with a Gaussian fitting of the line profile of the central 
spatial pixel (spaxel) and scaling using the standard point-source correction factors. 
The emission of all \water\ lines detected in the SSW band of the SPIRE-FTS, as well as most of the PACS lines, 
are unresolved  (diameter $ \leq 20\arcsec$, for the SPIRE lines, see Figure~\ref{lines_fwhm}, 
and $ \leq 10\arcsec$, for the PACS lines), indicating that water emission 
arises mainly from the CND region. Only one line, the p-\hho\,$2_{11}-2_{02}$ at 752.033~GHz 
was observed in the SLW band, with a beam size of $\sim33.4''$, and thus the emission of this
transition may have an additional contribution from the starburst ring.
%Two lines, p-\hho\,$2_{11}-2_{02}$ at 752.033~GHz and p-\hho\,$2_{02}-1_{11}$ 
%at 987.927~GHz, were observed in the SLW band, with a beam size of $\sim33''-37''$, and thus the emission of these 
%transitions may have an additional contribution from the starburst ring. 
However, our maps are not sensitive 
enough to disentangle the contribution of the starburst ring from the CND. 
Moreover, the PACS map of the o-\water\,$2_{12}-1_{01}$ line at 179~\mum\ (1669.905~GHz) shows 
extended emission (S. Hailey-Dunsheath private communication), 
most likely associated with the starburst ring. Thus, if there is some contribution from the 
extended emission in the central spaxel, the measured flux may overestimate the contribution of the nuclear flux.

\subsubsection{Excitation analysis}

In order to constrain the physical conditions of the water excitation, 
 we built a grid of models using the Large Velocity Gradient (LVG) model in plane parallel 
 geometry described in \citet{cec98} 
 varying the parameters in the following range: $n$(\hh)=10$^3$-10$^8$~\cmt, $T_{kin}=10-1500$~K, 
$N$(o-\water) and $N$(p-\water) from 10$^{11}$ to 10$^{17}$~\cmd.
The molecular data were taken from the BASECOL\footnote{http://basecol.obspm.fr} database \citep{dub06} 
and we used the most recent collisional rate coefficients with H$_2$ \citep{dub09, dan10, dan11}, 
calculated for temperatures between 5~K and 1500~K. The \water\ spectrum has
been computed considering 45 levels for the ortho and para water (i.e., considering the levels up to excitation temperatures of $\sim2000$~K). 
The two water forms are treated as independent species.
We used an ortho-to-para \hh\ ratio equal to 1 and an ortho-to-para \water\ ratio of 1. %The linewidth was fixed to 1~\kms.
The model includes the effects of the beam filling factor,
so it computes the $\chi^2$ for each column density minimizing with respect to the source size, temperature, and density. 
Our model does not include the effects of radiative pumping.

\begin{figure}[!ht]
 \includegraphics[width=\columnwidth]{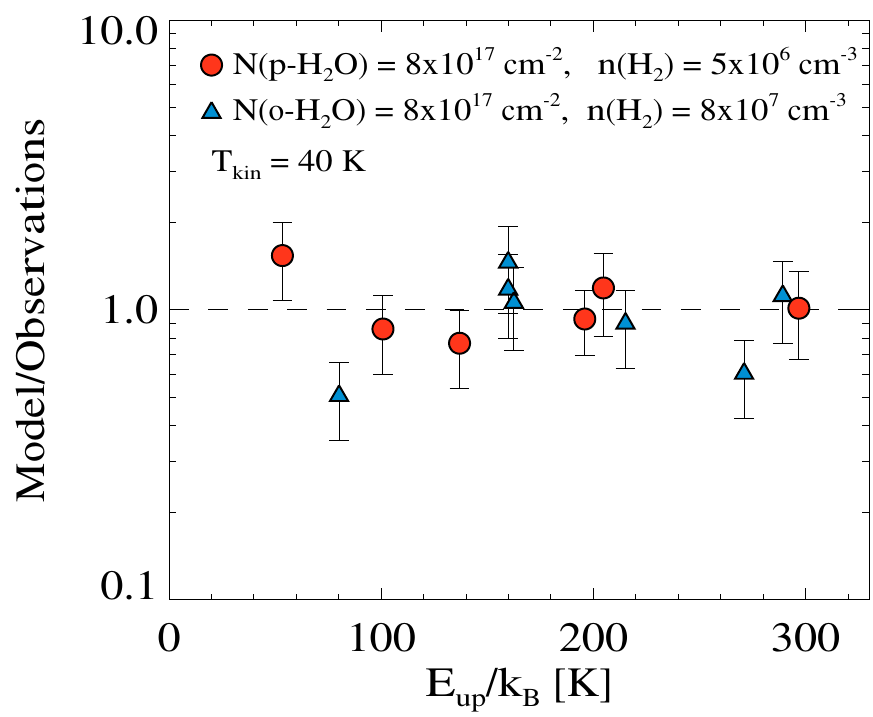}
\caption{Ratio between the best fit model predictions and the observations as a function of the upper level energy. 
Triangles show the o-\water\ transitions and circles show the p-\water\ 
transitions. 
The best fit models predict a column density of both ortho- and para-H$_{2}$O of $\sim$ 8 $\times$ 10$^{17}$ cm$^{-2}$ and
hydrogen desities in the range n(H$_{2}$) = 5 $\times$ 10$^6$ - 8 $\times$ 10$^7$ cm$^{-3}$. 
}
\label{lvgh2o}
\end{figure}

We fitted the o-\water\ and p-\water\ lines separately. We found that the observations of water lines are consistent with LVG model predictions for an 
emission size of $10''\pm2''$, a kinetic temperature of $T_{\mathrm{kin}}$ $\simeq$ 40$\pm$10~K, a density  n(H$_2$)  from  $4 \times 10^6$ 
to $8\times 10^7$ ~\cmt, 
and a value of the $N/\Delta\,v$ varying between $5 \times 10^{14}$ and $10^{16}$~\cmd.  
The column density of each water form can then be obtained once we know the linewidth, which we set to $\sim100$~\kms. 
This translates into a column density for both o-\water\ and p-\water\ around $5 \times 10^{16}-10^{18}$~\cmd. 
Figure~\ref{lvgh2o} shows the best fit model for the o-\water\ and p-\water\ lines. The results of 
the LVG model predictions are summarized in Table~\ref{h2oresults}. As can be seen in Fig.~\ref{lvgh2o}, 
the best fit model predictions can reproduce most of the line fluxes reasonably well.
 Only for two o-\water\  lines  the LVG model underestimates the
observed line flux. However, the line flux of the o-\water\  2$_{12}$-1$_{01}$  transition may be in fact overestimated 
due to some extended emission that may come from the starburst ring instead from the CND region 
(S. Hailey-Dunsheath private communication). 
As shown in previous sections and in \citet{hai12}, 
the CO emission from the CND is associated with several components at different temperatures and densities.
In the case of water we cannot rule out the presence of several components, but, due to the relatively low number of 
water lines, we did not attempt to fit the water lines with more than one component.

Our results suggest {\it a posteriori} that the excitation of the water lines is not strongly affected by radiative pumping.
Without the inclusion of radiative pumping, we can indeed fit all the
detected lines within a factor of two.  Moreover 
the water transitions in the SPIRE FTS spectrum that should be more affected by radiative pumping, 
besides that of  o-\water\  3$_{21}-3_{12}$, are those of o-\water\  5$_{23}-5_{14}$ and p-\water\  4$_{22}-4_{13}$ 
\citep[see, e.g.,][]{gon10}, which we do not detect,
consistent with the LVG model predictions.
On the other hand, we compare our findings with the results of Mrk\,231 \citep{gon10}, where all these three transitions 
have been detected and their fluxes can be reproduced only if radiative pumping is included. 
If we adopt the same line ratios of Mrk\,231 for NGC\,1068, we would expect to detect also these lines, originating from a higher level,  
at about a 10$\sigma$ level (assuming the 3$\sigma$ sensitivity in the SSW FTS range of  10$^{-17}$ W~m$^{-2}$).
We  conclude therefore that for NGC\,1068 collisions seem to be the dominant excitation mechanism for water and that radiative pumping
can be neglected.

\subsection{Molecular ions of H$_2$O$^+$, OH$^+$ and CH$^+$} 

The molecular ions of  \hdop, \ohp, and CH$^+$ have been detected in emission in NGC\,1068.
The 1115 GHz ground-state transition of  \hdop, as well as the 1033 GHz transition of \ohp\  and that of CH$^+$, 
have also been detected in emission in Mrk\,231 \citep{wer10}.
In contrast, these transitions have all been detected in absorption in Arp\,220 \citep{ran11} 
and two of them (\hdop\, at 1115 GHz and  \ohp\  at 1033 GHz) in absorption in M\,82 \citep{kam12}.

To obtain the column densities of H$_2$O$^+$, OH$^+$, and CH$^+$,  we assumed that the gas is in local 
thermodynamical equilibrium (LTE) and the emission is optically thin 
and originates in the CND region (diameter of 4$''$).  Assuming that all levels are populated according to the same 
excitation temperature, $T_{\mathrm{ex}}$, the column density is given by \citep[see, e.g.,][]{gol99}: 
\begin{equation}
N_{\mathrm{mol}}=\frac{4\pi\,F\,Q(T_{\mathrm{ex}})\exp(E_{\mathrm{up}}/k\,T_{\mathrm{ex}})}{h\nu\,A_{\mathrm{ul}}g_{\mathrm{u}}}
\label{coldens}
\end{equation}
where $F$ is the integrated line brightness (in units of erg\,cm$^{-2}$\,s$^{-1}$\,sr$^{-1}$), $\nu$ is the frequency of the transition, 
$E_{\mathrm{up}}$ the energy of the upper level, $Q$($T_{\mathrm{ex}}$) is the partition function at the excitation temperature $T_{\mathrm{ex}}$, 
and $h$ and $k$ are the Planck and Boltzmann constants, respectively.  We approximated the 
partition function of CH$^+$ and OH$^+$ to $Q_{\mathrm{rot}}\simeq\,kT_{\mathrm{ex}}/hB$, where $B$ is the rotational constant of the molecule. 
Einstein coefficients $A_{\mathrm{ul}}$, upper level degeneracy $g_{\mathrm{u}}$, and rotational constants have been obtained from the 
Cologne Database for Molecular Spectroscopy (CDMS; \citealt{mul01,mul05}).
For o-H$_2$O$^+$ the partition function has been 
obtained using the energy levels listed in the database. 
Both H$_2$O$^+$ and OH$^+$ present hyperfine structure, which is not resolved in the SPIRE FTS spectrum. 
Therefore, we  used $\sum_{i}\,A_{i\mathrm{l}}g_i$ to account for the several hyperfine components. 
We adopted an ortho-to-para ratio for H$_2$O$^+$ of 3. 

The estimated column densities are $N$(H$_2$O$^+$)=$(2.0-5.9)\times10^{13}$~\cmd, $N$(OH$^+$)=$(1.0-1.2)\times10^{13}$~\cmd, 
and $N$(CH$^+$)=$(5.4-7.5)\times10^{13}$~\cmd, for a range of excitation temperatures between 40 and 100~K. 
The uncertainty in the derived column densities is estimated to be around 50\%, which arises from the uncertainty in the flux calibration. 
The p-H$_3$O$^+$ $J_{\mathrm{K}}=3_{2}^{+}-2_{2}^{-}$ line at 364.7974 GHz has been detected in NGC 1068 using the JCMT 
\citep{aal11}. We recomputed the H$_3$O$^+$ column density using Eq.~\ref{coldens}, our adopted size of the CND region of $4''$,  
and the excitation temperature range of \Tex$\sim40-100$~K. We obtained $N$(p-\htop)$\simeq(1.8-0.7)\times10^{14}$~\cmd. 
Assuming an intermediate  ortho-to-para ratio of 1.5 \citep{van08},  
the total column density of \htop\ is $\sim(4.5-1.8)\times10^{14}$~\cmd. 

Considering the spectrum of Arp\,220 presented in \citet{ran11}, we notice that  in NGC\,1068, the ratio of
N(H$_3$O$^+$)/N(H$_2$O$^+$)$\sim$10, whereas this ratio is reversed in Arp\,220. Similarly, the inferred column densities 
indicate that N(H$_3$O$^+$)/N(OH$^+$) $\gg$ 1 in NGC\,1068, while in Arp\,220 the OH$^+$ column density is at least an order of
magnitude larger than that of H$_3$O$^+$. So the relative abundances of the molecular ions appear to be very different in the two galaxies,
suggesting some substantially different conditions.

We compared our estimated column densities with the predictions of the \citet{hol12} 
PDR models for different FUV incident fluxes and cosmic ray ionization rates. 
For the FUV radiation field produced by the AGN in the CND region (see Section~\ref{pdr_xdr}) the order of magnitude of the observed column densities 
of \hdop\ and \ohp\ are compatible with a low density medium (see Fig.~9 and Fig.~10 of \citet{hol12}). 
According to their models, the column density of  \htop\ is approximately constant with a value of $10^{14}$~\cmd, 
in agreement with our estimated value. 
Combining the observed value of the $N$(\ohp)/$N$(\hdop) ratio and the value of $N$(\ohp),  we estimated a range of values of 
log$_{10}[\zeta/n]$ between $-15.5$ and $-16.5$, where $\zeta$ is the cosmic ray 
ionization rate and $n$ is the hydrogen nucleus density, and we derived 
$N_{\mathrm{H}}>2\times10^{20}$~\cmd\ (see Figs.~10 to 13 of \citet{hol12}).  
Therefore PDR
models which include cosmic rays are in agreement with the observed column densities of these molecular ions.

\subsection{Hydrogen fluoride}
 
 \her~ spectroscopic observations 
have detected for the first time the J = 1-0 transition of hydrogen fluoride (HF) at 1232 GHz in the local universe,
and revealed the ubiquitous nature of this molecule in the interstellar medium (ISM) of
the Milky Way and of a few extragalactic sources, such as
the local AGN/ULIRGs  Mrk\,231 \citep{wer10},  
Arp\,220 \citep{ran11}, M\,82 \citep{kam12} and in the Coverleaf quasar at z=2.56 \citep{mon11}.

This transition is generally observed in absorption, as expected, due to its very
large Einstein coefficient, A$_{10}$ = 2.42 $\times$ 10$^{-2}$ s$^{-1}$. Only an extremely dense region, with a
strong radiation field, could generate enough excitation to yield an HF feature in emission \citep{neu09}.
 In NGC\,1068 it has been detected in emission (at a S/N $>$ 10), similarly to Mrk~231 and in contrast with Arp~220,
 where it has been detected in absorption.
 
 We estimated the column density of hydrogen fluoride using Equation~\ref{coldens} assuming that the 
 HF emission originated in the CND region (diameter $\sim4''$).  We adopted the 
 kinetic temperature $T_{\mathrm{kin}}\simeq100$~K derived for the CO gas associated with the CND region 
 (see Sec.3.2.1) as an upper limit to the excitation temperature. The values of the upper level degeneracy, $g_{\mathrm{up}}=3$,  
 and the rotational constant ($B$=616365~MHz) to calculate the partition function at a given temperature, 
 $Q_{\mathrm{rot}}\simeq\,kT_{\mathrm{ex}}/hB$, have been taken from the Jet Propulsion Laboratory (JPL) catalog \citep{pic98}. 
  For $T_{\mathrm{ex}}=40-100$~K, we found a column density of HF of $N$(HF)$\simeq7\times10^{12}$~\cmd. 
 This value is one order of magnitude lower than the $N$(HF) value found in M82 (Kamenetzky et al. 2012), 
 where hydrogen fluoride is seen in absorption. Calculations with RADEX 
 indicate that the optically thin approximation holds for a  wide range of physical conditions, 
 with $\tau\ll1$ for temperatures between $T_{\mathrm{kin}}=20-100$~K, densities $n$(H$_2$)$\simeq10^{3}-10^{6}$~\cmt, 
 and column densities of 10$^{11}$-10$^{14}$~\cmd.

\subsection{Atomic lines}

In this section, we analyze the origin of the three atomic fine-structure lines detected with the SPIRE FTS, 
namely,  the two lines from neutral carbon and the [NII]205~$\mu$m line. First we attempt to use the composite
photoionization model used by \citet{spi05}, which successfully fitted the overall UV to far-IR spectrum of NGC\,1068
reproducing the line fluxes within a factor 2 on average. This model was composed of an AGN component and a
starburst component, which included contribution from PDR clouds, as the integration was allowed to run
until the gas temperature in the cloud cooled down to T=50~K. We compare in the next section the predictions 
of the atomic lines detected by the SPIRE FTS using the same photoionization model reported in \citet{spi05}, 
which were not reported in that work.

\subsubsection{[NII] and [CI] emisson}

The observed [NII]205~$\mu$m line is roughly consistent with the photoionization model presented in  \citet{spi05}.
The predicted ratio [NII]205~$\mu$m/122~$\mu$m is $\sim$0.47, while the observed value is 18.8/30.5= 0.62$\pm$0.09,
using the ISO-LWS observation of the [NII]122~$\mu$m line \citep{spi05}. Taking into account that
the two lines have been observed by two different instruments, with different apertures, this result is satisfactory.
We can therefore conclude that the diffuse ionized emission traced by the [NII] ion is in agreement with being excited by the
starburst component in the SF-ring of NGC\,1068.
The observed [CI]369~$\mu$m/[CI]609~$\mu$m  line ratio is 1.65$\pm$0.30, while the ratio predicted by the photoionization 
model is about 3.8, indicating that the emission of the line at 609~$\mu$m could be more extended than the FTS beam. 

On the other hand, the intensity of the [CI] lines predicted by the photoionization model are weaker by more than an order of 
magnitude compared to the observed values relatively to the [CII]158~$\mu$m emission, as measured in \citet{spi05}.
The predicted ratio of  [CI]369~$\mu$m/[CII]158~$\mu$m is in fact $\sim$ 10$^{-3}$ and the predicted [CI]609/[CII]158~$\mu$m~$\mu$m
is $\sim$ 3 $\times$ 10$^{-4}$, while the observed ratios are 9.7 $\times$ 10$^{-3}$ and 5.9 $\times$ 10$^{-3}$, respectively.
This discrepancy between data and models could be due to the poor ability of current models to reproduce the [CI] emission line intensities
\citep[see, e.g.,][]{rol07}.
 
\subsubsection{Using Neutral Carbon emission to estimate the gas temperature}

We derive here the kinetic temperature of the molecular gas from the ratio of the intensities of the neutral Carbon lines ([CI] at 369 and 609~$\mu$m), 
when local thermodynamic equilibrium (LTE) is assumed.
The Boltzmann equation can be written as
\begin{equation}
T_k = {-E_{21} \over k} \left[ ln \left({g_1 n_2\over g_2 n_1}\right) \right]^{-1} 
\end{equation}
where $T_k$ is the kinetic temperature of the gas, $E_{21}=E_2-E_1$, $n_1$ and $n_2$ are the level populations and $g_1$ and $g_2$ 
their statistical weights. The integrated flux in a line is simply $F_{ij}\propto A_{ij} n_i \nu_{ij}$.
 Therefore the kinetic temperature can be written as
\begin{equation}
T_k = {-E_{21} \over k} \left[ ln \left({g_1 \over g_2} \cdot {{A_{10} \nu_{10} F_{21}}\over {A_{21} \nu_{21} F_{10}}}\right) \right]^{-1} 
\end{equation}

By substituting the observed values of the fluxes of the [CI] lines at 492.16 and 609.34 GHz 
of Table \ref{tab:lines} and propagating the uncertainties, we obtain a kinetic temperature of the gas of T$_k$ = 22.5 $\pm$ 2.4 K.
The kinetic temperature derived from the two [CI] lines is much lower than for either the CND or extended starburst ring
components, which again argues that the CI emission is much more extended than either of these.

\subsection{Comparing the derived masses with previous work}

We discuss here our results in the context of the various molecular observations of NGC\,1068
in the literature.
In particular our radiation transfer models indicate the presence of two major components 
responsible for the CO (J$_{up}$ $\leq$ 13) excitation: the first one is a compact component, 
with a diameter of $\sim$ 4$\arcsec$ (300 pc),  associated with the CND with a density of 
n(H$_{2}$) $\sim$ 4$\times$10$^{4}$~cm$^{-3}$, kinetic temperature of T$\sim$90 K and mass 
M(H$_{2}$) $\sim$ 2.4$\times$10$^{7}$~M$_{\sun}$, and the second is an extended component,
with a diameter of $\sim$ 40$\arcsec$ (3 kpc), associated with the SF-ring,  
with a density of n(H$_{2}$) $\sim$ 7$\times$10$^{2}$~cm$^{-3}$, kinetic temperature of T$\sim$116 K and mass 
M(H$_{2}$) $\sim$ 3.5$\times$10$^{8}$~M$_{\sun}$.

\citet{sch00} have used interferometric CO (1-0) and (2-1) observations to derive a mass of the CND (``ring" in CO (2-1), in their terminology,
of $\sim$200~pc  in radius) of
M(H$_{2}$) $\sim$ 5$\times$10$^{7}$~M$_{\sun}$, while they measure a mass of the "spiral arms" (our SF-ring) of 
M(H$_{2}$) $\sim$ 6.8$\times$10$^{8}$~M$_{\sun}$. These two mass estimates are a factor 2 
times larger than our
estimates; however,  this discrepancy is fully justified by the difference in the gas excitation, as the CO (1-0) and (2-1) lines are mapping
lower temperature gas.

We then compare our findings with the temperature of T$\geq$145~K and mass derived from the H$_{2}$ pure rotational transitions
of \citet{rig02}; from the detection of the S(1) line \citep{lut00} a warm gas mass of M(H$_{2}$) $\sim$ 10$^{8}$~M$_{\sun}$
is derived in the SWS beam of 14$\arcsec$ $\times$ 27$\arcsec$ \citep{deg96}. This is consistent with our findings, given the
difference in beam size between the two spectrometers.

\subsection{Observability of submillimetre lines at high redshift }

\begin{figure*}[!ht]
 \includegraphics[width=\columnwidth]{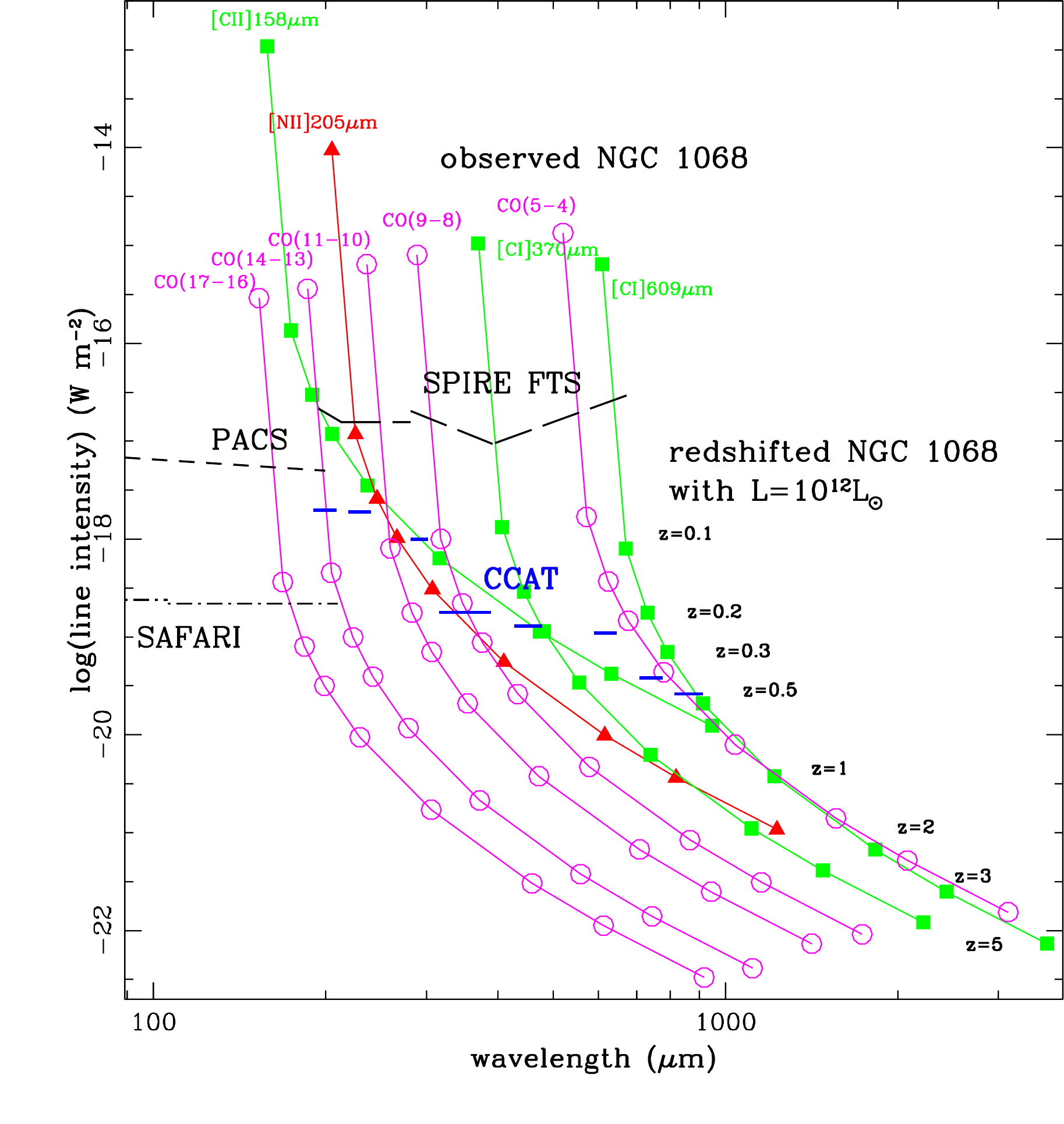}
 \includegraphics[width=\columnwidth]{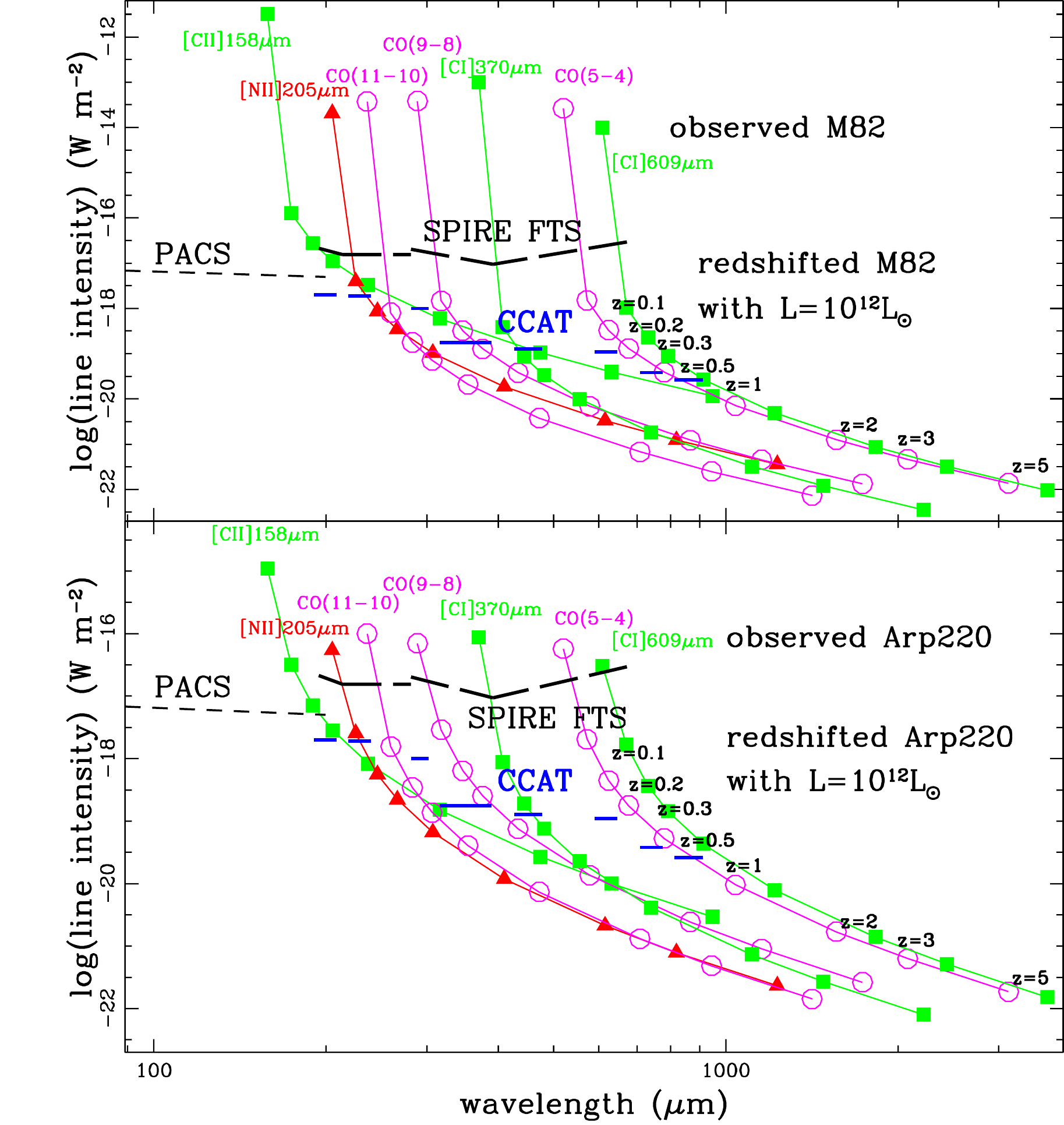}
 \caption{{\it Left panel:} Comparison between the observed far-IR/submillimetre lines of NGC\,1068 with the predictions of the 
same lines as a function of redshift of a similar galaxy with an intrinsic luminosity of L=10$^{12}$L$_{\sun}$.
The stellar/HII region lines are shown as filled triangles (in red), the low-ionization/PDR lines are shown as filled squares (in green),
while the molecular lines are shown as open circles (in magenta).
For comparison, we overplot the 5$\sigma$ (in 1 hr) sensitivity threshold of \her -PACS, \her -SPIRE, SPICA-SAFARI and CCAT.
See \citet{spi12} for more details on the assumed sensitivities of the future SAFARI and CCAT instruments.
{\it Right panel:}  Same diagram for other two templates: the starburst galaxy M82 \citep{kam12} and the ULIRG Arp220 \citep{ran11}.
}
\label{ccat}
\end{figure*}

Using NGC\,1068 as a template to predict the submillimetre spectrum of higher redshift galaxies, we have estimated the expected
line fluxes as a function of redshift, rescaling the total infrared luminosity of NGC\,1068 to the value of L$_{IR}$=10$^{12}$L$_{\sun}$. 
We show in Figure ~\ref{ccat}(a) the observed fluxes of atomic and molecular lines of NGC\,1068 from the ISO-LWS spectrometer \citep{spi05}, from the 
PACS spectrometer \citep{hai12}, and those from this work, and compare them with the predicted fluxes at redshift of z=0.1, 0.2, 0.3, 0.5 and
1, 2, 3, 5. We also show in the figure the 5$\sigma$ 1 hour sensitivities of future observing facilities, such as CCAT and SPICA-SAFARI and compare
them with the sensitivities of the \her~ spectrometers. We refer to \citet{spi12} and references therein for a brief description of SPICA and CCAT
and for the details of the expected sensitivities of the foreseen spectrometers at their focal planes. We also show in Figure ~\ref{ccat}(b), 
for comparison with NGC\,1068, the observed and predicted submillimetre spectra of other important local template galaxies, 
the prototypical starburst galaxy M82 and the ULIRG Arp220, that have been observed with the SPIRE spectrometer \citep[][respectively]{kam12,ran11}.

It is clear from Figure ~\ref{ccat} that the intermediate to high-J CO lines, as well as the two [CI] lines, are weaker by 1-2 orders
of magnitude compared to the brightest far-IR fine structure lines of [CII] {\it and [NII]}. 
However, their powerful diagnostic potential, in terms of detecting through XDR regions the effect of an AGN in the host galaxy, 
 combined with their long wavelengths, makes these lines very attractive for high redshift 
spectroscopic cosmological surveys from ground-based telescopes.
The expected sensitivity of future 
spectrometers at the focal plane of large submillimetre telescopes, such as CCAT, will be able to detect an object like NGC\,1068, M82, or Arp220 with
a luminosity of L=10$^{12}$L$_{\sun}$ at a redshift of z=0.2-0.5. 

\section{Summary and conclusions}

We summarize here the results of this work.
The first complete submillimetre (190--670~$\mu$m) spectrum of the Seyfert type 2 galaxy NGC\,1068 reveals the
full sequence of CO pure rotational lines from J$_{up}$=4 to J$_{up}$=13. The radiation transfer analysis of these lines
shows the presence of two physically distinct components: the first one originating from the circum-nuclear disk (CND)
of few arcseconds in diameter ($\sim$ 4$\arcsec$) and the second one excited in the star forming ring (SF-ring) with a diameter 
ten times larger ($\sim$ 40$\arcsec$). These results indicate a kinetic temperature of CO  of $T_{\rm kin}$=100~K and  127~K, a
gas density of $n({\rm H_2})$=$10^{4.5}$ and $10^{2.9}$ cm$^{-3}$ and a derived molecular hydrogen mass of
M(H$_{2}$) $\sim$ 2.4$\times$10$^{7}$~M$_{\sun}$ and M(H$_{2}$) $\sim$ 3.5$\times$10$^{8}$~M$_{\sun}$,
 for the compact and extended regions, respectively.

The comparison of the observed CO line intensities with predictions of photodissociation (PDR) and X-ray dominated regions (XDR) models shows
that the circum-nuclear disk emission can be modeled equally well by both types of models, while the CO lines in the star-forming ring can be
modeled by a photodissociation region only. However some observational constraints, such as the total CO surface brightness and the required
radiation field, indicate that the most plausible explanation for the CO excitation of the CND is an XDR with density of n(H$_2$) $\sim$ 10$^4$ cm$^{-3}$
and X-ray flux of 9 erg~s$^{-1}$~cm$^{-2}$, consistent with the AGN illumination. In contrast, the excitation of the SF-ring component is due to 
PDR emission originating from the young stars/HII regions in the spiral arms. 

The water lines that we have detected with SPIRE, together with those observed by PACS (S. Hailey-Dunsheath 2012, private comm.),
have been modeled with an LVG model 
to constrain the physical conditions of the water excitation. We have found that the kinetic temperature is T$_{kin}$=40~K, 
the molecular hydrogen density is $n(H_2)=4 \times 10^6 - 8\times 10^7$ and the column density is of  order 
 N(H$_2$O)=8 $\times 10^{17}$ cm$^{-2}$ for both water forms.

The computed column densities of the molecular ions detected (\hdop\ and \ohp\ ) are in agreement with PDR models that include
cosmic ray ionization. 

The fundamental rotational transition of HF has been detected in emission in NGC\,1068 and we infer a column density of 
N(HF)$\sim 7.5\times 10^{12}$ cm$^{-2}$.

For the two [CI] transitions, we derived a kinetic temperature of 22.5$\pm$2.4 K in LTE approximation, which is
much lower than the temperatures traced from the intermediate-J CO molecular gas.

The molecular masses that we derived from our analysis are in good agreement with the masses estimated from
both CO interferometric measurements of low-J lines and mid-infrared H$_2$ emission lines.

Finally we show that  the intermediate-J CO and [CI] lines in galaxies with L$\sim10^{12}L_{\sun}$
can be observed from planned and future ground-based and
space telescopes up to redshift of z $\sim$ 0.5, making their diagnostic power an important tool to study galaxy
evolution at intermediate redshift.
 
\acknowledgments

We thank Nicola Sacchi for having provided a preliminary reduction and analysis of the SPIRE FTS
spectrum of NGC\,1068, Steven Hailey-Dunsheath for having provided to us PACS data before publication 
and Scige Liu for assistance in the SPIRE data reduction. We also thank the anonymous referee for his/her
careful analysis of the paper that helped improving it.

SPIRE has been developed by a consortium of institutes led
by Cardiff Univ. (UK) and including: Univ. Lethbridge (Canada);
NAOC (China); CEA, LAM (France); IFSI, Univ. Padua (Italy);
IAC (Spain); Stockholm Observatory (Sweden); Imperial College
London, RAL, UCL-MSSL, UKATC, Univ. Sussex (UK); and Caltech,
JPL, NHSC, Univ. Colorado (USA). This development has been
supported by national funding agencies: CSA (Canada); NAOC
(China); CEA, CNES, CNRS (France); ASI (Italy); MCINN (Spain);
SNSB (Sweden); STFC, UKSA (UK); and NASA (USA).
This work has been funded in Italy from ASI (Italian Space Agency)
under contract I/005/11/0.
The research of C.D.W. is supported by grants from NSERC (Canada).

%\clearpage

%\bibliographystyle{apj}
%\bibliography{n1068_7.7.2012}

\clearpage

%\begin{landscape}
\begin{deluxetable}{llcccccccc}%[!h]
\tabletypesize{\scriptsize}
\tablecaption{ Fluxes of the molecular and atomic lines detected in the SPIRE FTS spectrum of NGC\,1068 and compared to literature data
\label{tab:lines} } 
\tablehead{ \colhead{species}& \colhead{line id.}  & \colhead{rest. freq.} & \colhead{$E_{\mathrm{up}}/\kappa_{B}$} & \colhead{Beam}  & \colhead{Flux} & \colhead{1$\sigma$ Stat. Uncert.}   & \colhead{1$\sigma$ Tot. Uncert.}   &\colhead{Ref.} \\
%\\
               &  	& 	            &         &   FWHM & &&&\\
\cline{6-8} \\
 \colhead{}   &   \colhead{}                  & \colhead{(GHz)} & \colhead{(K)}  & \colhead{($\arcsec$)}     &  \multicolumn{3}{c}{ (10$^{-17}$ W m$^{-2}$) } &  \colhead{}  \\}
\startdata
\multicolumn{2}{c}{compact  source}  &                &  & && && & \\
\hline
               &  	& 	            &         & &  &      &       &    \\
 $^{12}$CO     &   J = 1-0  &    \phn115.271 &  5.53    & 1 $\times$ 0.8 &   .050   & .001   & 0.01 & 1  \\%$^{12}$CO   
   &   J = 2-1                 &    \phn230.538        & 16.60   & 1 $\times$ 0.8 &  .360   & .003   & 0.07 & 1  \\ %$^{12}$CO  
    &   J = 3-2                 &    \phn345.796       &  33.19  & 1 $\times$ 0.8 &  2.36   & 0.34   & 0.50    &1  \\
     &   J = 9-8                  &   1036.912 & 248.88            & 18.7 &15.41   &  0.15 &  1.55  & 2  \\ % $^{12}$CO
     &   J = 10-9                &   1151.985 & 304.16            & 17.1 &14.32$^{\mathrm{a}}$ &  0.23 & 1.45  & 2  \\ % $^{12}$CO
     &   J = 11-10              &   1267.014 & 364.97            & 17.6 &12.80   &  0.15 & 1.29  & 2  \\% $^{12}$CO
     &   J = 12-11              &   1381.995 & 431.29            & 16.9 &10.97   &  0.18 & 1.11  & 2  \\% $^{12}$CO
     &   J = 13-12              &   1496.923 & 503.13            & 16.8  & 6.11   &   0.27 & 0.67  & 2  \\
                                                             &  	      & 	        &         &         &           &       &    \\
o-\hho & $3_{12}-3_{03}$\phs       &1097.365  & 215.1    & 18.7            &4.15 & 0.03 & 0.42 & 2\\ %ok
&$3_{21}-3_{12}$\phs 	   	     &1162.912  & 271.0    & 17.1            &5.80 & 0.14 & 0.60 & 2 \\ %ok
&  $2_{21}-2_{12}$                          &1661.008   & 159.8   & 12.7 	   & 4.8   & \nodata & 1.6 & 3\\
& $2_{12}-1_{01}$                           &1669.905    & 80.1     & 12.7	   & 14.3 & \nodata & 4.3 & 3\\
& $3_{03}-2_{12}$                           &1716.769    & 162.5   & 12.3	   & 6.9   & \nodata & 2.2 & 3\\
& $4_{14}-3_{03}$                           &2640.474    & 289.2   & 8.0	            & 5.1$^{\mathrm{b}}$ & \nodata & 1.6 & 3\\
& $2_{21}-1_{10}$                           &2773.978    & 159.8   & 7.6	            & 6.5 & \nodata & 2.1 & 3\\
p-\hho & $2_{11}-2_{02}$\phs     &\phn752.033 & 136.9  & 33.4           & 3.42 & 0.07 & 0.35 & 2 \\ %ok	
& $2_{02}-1_{11}$\phs 		  &\phn987.927  & 100.8  &  19.8	  & 6.09 & 0.11 & 0.62 & 2	\\  %ok
& $1_{11}-0_{00}$\phs 		  &1113.343    	  & 53.4    &  18.7	  & 3.49 & 0.05 & 0.35 & 2 \\ %ok
& $2_{20}-2_{11}$\phs 		  &1228.789        & 195.9  &  17.6         & 5.30 & 0.09 & 0.54 & 2\\ %ok
& $3_{22}-3_{13}$                        &1919.359         & 296.8  &   9.8  	  &2.4 & \nodata & 0.8 & 3 \\
& $3_{13}-2_{02}$                        &2164.132         & 204.7  &  11.5	  &6.9 & \nodata & 2.2 & 3 \\
	                                                &  			   & 	        &                   &	  &                &	  &    \\
 o-\hho$^+$  &  $1_{11}-0_{00}$ J=3/2-1/2 &   1115.204   & 53.5      & 18.7       & 2.13   &   0.05  &  0.22  &  2 \\ %ok
 o-\hho$^+$  &  $1_{11}-0_{00}$ J=1/2-1/2  &   1139.561  &  53.5    & 17.1       & 1.98   &   0.09  &  0.22  &  2 \\ %ok
 OH$^+$       &  1$_{12}-0_{12}$                     &  1033.118   &  49.6    & 18.7       & 4.34    &   0.16   & 0.46   &  2  \\ %ok
 CH$^+$       &  J = 1-0                                      &  835.079      &  39.6    & 35.0       & 3.16    &   0.08   &  0.33  &  2  \\ %ok
  HF                 &   J = 1-0                                   &   1232.476   & 59.1     &  17.6      & 2.16    &   0.09  &  0.23 & 2  \\ %ok
                          &  			     & 	                    &               &            &             &              &            &       \\
{\rm [NII]}     &  $^3P_1 - ^3P_0$     &   1461.134 & 6.77      & 16.8 &  54.97  &   1.42     &    5.67    & 2 \\ 
                          &  			     & 	                    &               &            &             &              &            &       \\
\hline											
\multicolumn{2}{c}{extended source}                           &    & & & & & && \\
\hline											
               &  			     & 	            &         &       &  &     &  \\
 $^{12}$CO     &   J = 2-1                 &    230.538   & 16.60  & 31.0   &  6.41    & .015    &  1.3   &  4  \\% $^{12}$CO
                          &   J = 3-2                 &    345.796   &  33.19 & 43.4   &  19.4    & \nodata          &    3.4   &  5  \\% $^{12}$CO
                          &   J = 4-3                 &    461.041   &  55.32 & 43.4   &  21.69    &  0.66    &  2.73   &  6 \\% $^{12}$CO
                          &   J = 5-4                 &    576.268   &  82.97 & 43.4   &  26.80    &  0.43    &  3.30   &  6 \\% $^{12}$CO
                          &   J = 6-5                 &    691.473   &  116.16 & 43.4 &  24.73    &  0.24    &  3.02  &  6 \\% $^{12}$CO
                          &   J = 7-6                 &    806.652   & 154.87 & 43.4  &  21.06    &  1.30    &  2.88  &  6 \\ % $^{12}$CO
                          &   J = 8-7                 &    921.800   & 199.11 & 43.4  &  19.78    &  0.64    &  2.50  &  6  \\
                          &  			     & 	                    &               &            &             &              &            &       \\
{\rm [CI]}          & $^3P_1 - ^3P_0$ &  492.161     &  23.62  & 43.4   &  12.81   & 0.13     &  1.57  & 6  \\ %{\rm [CI]}
                         & $^3P_2 - ^3P_1$ &  809.342     & 62.46   & 43.4   &  21.08   & 1.20        &  2.84  & 6 \\
                          &  			     & 	                    &               &            &             &              &            &       \\
{\rm [NII]}     &  $^3P_1 - ^3P_0$     &   1461.134 & 6.77      & 43.4 &  187.9  &   8.2     &     24.3        & 6 \\ 
\enddata
\tablecomments{
$^{\mathrm{a}}$ This line is possibly contaminated by the o-H$_2$O   $3_{12}-2_{21}$ at 1153.127 GHz.
However, the line fitting results indicate that the CO line is dominant; therefore we assign the whole flux to the CO
line; 
$^{\mathrm{b}}$ This line is blended with the CO\,(23--22) transition, so to obtain the flux of water we 
subtracted the flux of CO\,(23--22) predicted by the model presented by \citet{hai12}.}
\tablerefs{
(1): Interferometric observations from \citet{kri11}, adding together the emission from the E-knot and the W-knot 
and using the conversion: $f ({\rm W~m^{-2}}) = 3.33 \times 10^{-23} \nu_{GHz} S_{CO}({\rm Jy~km~s^{-1}})$; 
(2): This work, SPIRE-FTS deep pointed spectrum; 
(3): S. Hailey-Dunsheath, 2012, private communication;
(4): From CSO Z-Spec observations of \citet{kam11}, using the same conversion as for (1); 
(5): C. Wilson, 2012, private communication; 
(6): This work, SPIRE-FTS line maps}
\end{deluxetable}
%\end{landscape}

\begin{deluxetable}{ccc}%[h!]
   \tablewidth{0pt}
   \tablecaption{RADEX Model Parameters and Ranges\label{table:radex}}
   \tablehead{
   \colhead{Parameter} & \colhead{Range} & \colhead{\# of Points}
	}
   \startdata
$T_{kin}$ [K] & $10^{1.7} - 10^{3.5}$ & 90\\
$n(H_2) $ [cm$^{-3}$] & $10^{1.0} - 10^{5.0}$ &  90\\
$N_{\rm CO}$ [cm$^{-2}$] & $10^{16} - 10^{20}$ &  90\\
$\Delta V$ [km s$^{-1}$] & 1.0 &  fixed
%$T_{background}$ [K] & 2.73 &  fixed
   \enddata
\tablecomments{All parameters are sampled evenly in log space.}
\end{deluxetable}

\begin{deluxetable}{ccccc}
\footnotesize
\tablewidth{0pt}
\tablecaption{Multi-Component Radiative Transfer 
Results\label{table:radexresults}}
\tablehead{
\colhead{Parameter} & \colhead{Median} & \colhead{1$\sigma$ Range} & 
\colhead{1D Max} & \colhead{4D Max} \\
}
\startdata
compact source & &&&\\
\hline
%\startdata
&&&&\\
$T_{\rm kin}$ [K]                   &  92                  &  74 $-$  120                      &  87                            &  100              \\
$n({\rm H_2})$ [cm$^{-3}$] & $10^{4.6}$ & $10^{4.4} - 10^{4.8}$ & $10^{4.6}$            & $10^{4.5}$ \\
$N_{CO}$ [cm$^{-2}$]         & $10^{19.5}$ & $10^{19.3} - 10^{19.7}$ & $10^{19.5}$  & $10^{19.6}$ \\
$\Phi_{\rm A}$                       & $10^{-0.9}$  & $10^{-1.0} - 10^{-0.9}$   & $10^{-0.9}$   & $10^{-0.9}$  \\
%\hline
$P$ [K cm$^{-2}$]                & $10^{ 6.6}$   & $10^{6.5} - 10^{6.8}$      & $10^{6.6}$      & $10^{6.6}$ \\
$\langle N_{\rm CO} \rangle$ [cm$^{-2}$] & $10^{18.6}$ & $10^{18.4} - 10^{18.7}$ & $10^{18.6}$ & $10^{18.6}$\\
M(H$_2$) [M$_\odot$]           & $10^{7.4}$    & $10^{7.2} - 10^{7.5}$    & $10^{7.4}$     & $10^{ 7.4}$ \\
&&&&\\
\hline
extended source & & & & \\
\hline
&&&&\\
$T_{\rm kin}$ [K]                    &   116               &   80 $-$  166                           &   121             &   127              \\
$n({\rm H_2})$ [cm$^{-3}$] & $10^{2.8}$   & $10^{2.3} - 10^{3.3}$      & $10^{2.9}$ & $10^{2.9}$  \\
$N_{CO}$ [cm$^{-2}$]         & $10^{18.8}$ & $10^{18.3} - 10^{19.3}$ & $10^{18.7}$ & $10^{18.7}$ \\
$\Phi_{\rm A}$                       & $10^{-1.2}$  & $10^{-1.3} - 10^{-1.0}$   & $10^{-1.2}$ & $10^{-1.1}$  \\
%\hline
$P$ [K cm$^{-2}$]                & $10^{ 4.7}$      & $10^{ 4.3} - 10^{ 5.2}$    & $10^{4.4}$ & $10^{4.4}$ \\
$\langle N_{\rm CO} \rangle$ [cm$^{-2}$] & $10^{17.7}$ & $10^{17.2} - 10^{18.2}$ & $10^{17.6}$ & $10^{17.6}$ \\
M(H$_2$) [M$_\odot$]           & $10^{8.5}$    & $10^{ 8.0} - 10^{ 9.0}$    & $10^{ 8.5}$ & $10^{ 8.5}$ \\
&&&&\\
%\hline
\enddata
\tablecomments{1D max refers to the maximum value of the integrated 
parameter distribution.
4D max refers to the value of that parameter at the best fit solution.}
\end{deluxetable}

\begin{deluxetable}{cc}%[hb!]
   \tablewidth{0pt}
   \tablecaption{The best fit model parameters of \water \label{h2oresults}}
  \tablehead{
  \colhead{Parameter} & \colhead{Value}
	}
   \startdata
Temperature &$40\pm10$~K  \\
density     & $4 \times 10^6 - 8 \times 10^7$~\cmt\  \\
diameter	    &10$''\pm2''$ \\
$N$(o-\water)& $2 \times 10^{17} - 10^{18} $~\cmd\ \\
$N$(p-\water)& $5 \times 10^{16} - 10^{18} $~\cmd\ \\
   \enddata
\end{deluxetable}

\end{document}